\documentclass[a4paper,11pt]{article}
\usepackage{jheppub} 
\usepackage{lineno}

\usepackage{epsfig} 
\usepackage{amsmath} 
\usepackage{graphicx}
\usepackage{color}
\usepackage{float}
\newcommand{\ba}{\begin{array}}
\newcommand{\ea}{\end{array}}
\newcommand{\bea}{\begin{eqnarray}}
\newcommand{\eea}{\end{eqnarray}}

\newcommand{\be}{\begin{equation}}
\newcommand{\ee}{\end{equation}}

\def\bra#1{\left\langle #1\right|}
\def\ket#1{\left| #1\right\rangle}

\def\bs{B_s^0}

\def\bsbar{{\bar B}_s^0}

\title{\boldmath Probing quantum decoherence at Belle II and LHCb}

\author[a]{Ashutosh Kumar Alok,}
\author[a]{Subhashish Banerjee,}
\author[a]{Neetu Raj Singh Chundawat} 
\author[b]{and S. Uma Sankar}

\affiliation[a]{Indian Institute of Technology Jodhpur, Jodhpur 342037, India}
\affiliation[b]{Indian  Institute  of  Technology  Bombay,  Mumbai  - 400076, India}

\emailAdd{
akalok@iitj.ac.in}
\emailAdd{subhashish@iitj.ac.in}
\emailAdd{chundawat.1@iitj.ac.in }
\emailAdd{uma@phy.iitb.ac.in}

\abstract{With the advent of Belle II and the LHCb upgrade, the precision measurements of various B-Physics observables are on cards. This holds significant potential for delving into physics beyond the standard model of electroweak interactions. These measurements can also serve as means to establish limits on phenomena occurring at much finer length scales, such as quantum decoherence, which may arise due to potential discreteness in space-time or non-trivial topological effects. In this work, we set up the formalism to investigate the impact of quantum decoherence on several potential observables in $B$ meson systems. The approach employs the trace-preserving Kraus operator formalism, extending unitary evolution to non-unitary dynamics while maintaining complete positivity.  In this formalism, the decoherence effects are parametrized in terms of a single parameter. Through the analysis of purely leptonic, semileptonic, and non-leptonic decays of $B$ mesons, we identify observables that could, in principle, be influenced by decoherence. The theoretical expressions are provided without neglecting the impact of decay width difference ($\Delta \Gamma$) and $CP$ violation in mixing. Considering that many of these observables can be measured with high precision using the abundant data collected by LHCb and Belle II, our formalism can be applied to establish constraints on the decoherence parameter through multiple decay channels. This offers an alternative set-up for such studies, which, at present, are predominantly conducted in the neutrino sector.}

\begin{document}
\maketitle
\flushbottom

\section{Introduction}

Flavor physics, particularly the analysis of $B$ meson decays, has not only subjected the Standard Model (SM) of electroweak interactions to rigorous tests but has also imposed constraints on various extensions beyond the SM. With the absence of direct observations of new physics in high-energy experiments like ATLAS and CMS, the significance of flavor physics experiments, such as LHCb and Belle II, becomes more pronounced. These experimental setups have the capacity to explore new physics realms, reaching scales well beyond the capabilities of energy frontier experiments, by probing virtual particle production in quantum loops. The versatility of flavor systems also extends to the investigation of physics emerging from much finer length scales, including phenomena like quantum decoherence and Lorentz violation \cite{Ellis:1983jz,Ellis:1992dz,Huet:1994kr}. In this study, we examine the potential effects of quantum decoherence in B meson systems.

In a series of seminal works, S. Hawking provided a groundbreaking insight into the thermal radiation emitted by black holes, establishing a significant connection between quantum mechanics and the behavior of these celestial entities \cite{Hawking:1974rv,Hawking:1975vcx,Hawking:1982dj}. Specifically, in ref. \cite{Hawking:1975vcx}, it was pointed out that black holes seemed to necessitate a mixed statistical description. This was intuitively attributed to the loss of information across the event horizon. In ref. \cite{Hawking:1982dj}, it was demonstrated that a pure quantum state 
undergoes an intriguing transformation into a mixed state in the presence of a microscopic event horizon. This transition gives rise to a phenomenon known as quantum decoherence, a departure from the standard principles of quantum mechanics. Quantum decoherence, typically forbidden in the framework of standard quantum mechanics due to its violation of unitarity, introduces a unique departure from the expected behavior of quantum systems. The violation of unitarity implies a loss of information about the system, leading to the emergence of mixed states from initially pure quantum states. The implications of this departure from unitarity extend to the foundational principles of theoretical physics, notably the Charge-Parity-Time (CPT) theorem. In flat space-times, the CPT theorem holds as long as the underlying theory adheres to principles of locality, unitarity, and Lorentz invariance \cite{Streater:1989vi}. Consequently, the occurrence of quantum decoherence introduces a potential breach in CPT symmetry by contravening the unitarity condition.

The distinctive topological features inherent in this environment may give rise to the spontaneous creation of quantum black holes characterized by event horizons with radii comparable to the Planck length. This continuous cycle of creation and evaporation, orchestrated by Hawking radiation, imparts a dynamic and frothy nature to the fabric of spacetime \cite{Hawking:1979hw,Hawking:1979pi}.  As particles navigate the vicinity of these quantum black holes, a noteworthy interaction unfolds. The environment surrounding these singularities absorbs specific degrees of freedom from passing particles. This interaction triggers a transformative process in the states of these particles, transitioning them from pure to mixed states. 

Quantum decoherence manifests in a specific category of string theories, namely non-critical string theories \cite{noncritical,Ellis:1997jw,Mavromatos:2004sz}. In the realm of quantum loop gravity, space-time is characterized by discreteness, and a study demonstrated that this discrete nature of space-time is conducive to the occurrence of quantum decoherence \cite{Gambini:2003pv}. Apart from the decoherence arising from space-time foam effects, an additional source of decoherence stems from dark energy. 
The acceleration of the Universe is propelled by dark energy, prompting the evolution of the cosmos into a de Sitter universe characterized by exponential expansion. This transformation implies the presence of a cosmological horizon, and this, in turn, introduces the possibility of decoherence due to the challenge of defining super-scattering operators that connect the initial and final states. This decoherence of cosmological origin was shown to be intimately related to quantum gravity effects in ref. \cite{Mavromatos:2003hr}.

The dynamics governing the entire system, encompassing both the system and its surrounding environment, adhere to unitary principles, preserving quantum coherence and overall system unitarity. However, when we focus solely on the system and trace over the environmental degrees of freedom, the system's evolution becomes non-unitary. Regardless of the origin of the environment at the microscopic level, its influence on the system can be effectively addressed by employing the principles of open quantum systems \cite{Lindblad:1975ef,Gorini:1975nb,Gorini:1976cm,Matthews:1961vr,kraus,alicki,breuer,Caban:2007je,Caban:2005ue,weiss,banerjee}. This approach provides a systematic framework for incorporating the impact of external environments, including decoherence and dissipation, into the dynamics of quantum systems.

In this study, we establish a formalism aimed at examining decoherence effects within the $B$ meson system, focusing on observables of potential significance in experiments such as LHCb and Belle II. These observables pertain to a range of decay processes, encompassing purely leptonic, semileptonic, and non-leptonic decays of B mesons. The evolution of the density matrix is governed by Kraus operators, extending the unitary evolution to a non-unitary framework while maintaining completely positive dynamics. These Kraus operators are characterized by a single parameter that models decoherence, in addition to other parameters that are specific to the meson system.  Therefore our approach is phenomenological in nature, which means that the effective description is relatively independent of the intricate details of the actual dynamics governing the interaction between the subsystem and its environment. Also, we present expressions without neglecting decay width difference $\Delta \Gamma$ and $CP$ violation in mixing. 
The inclusion or exclusion of these effects in an analysis depends on the theoretical and experimental precision associated with a particular decay mode. Furthermore, in this formalism, the resulting density matrices representing the meson evolution are trace-preserving. This approach provides a unified framework to investigate both particle physics observables as well as quantum correlations in meson system as the measures of quantum correlations are defined for trace-preserving density matrix, see for e.g., \cite{Horodecki:1996rv,Blasone:2007wp,Banerjee:2014vga,Naikoo:2018vug,Dixit:2018gjc,Alok:2015iua}.

In ref. \cite{Alok:2015iua}, it was shown for the first time that the determination of observables such as $B-\bar{B}$ mixing parameter $\Delta m$ and $\Delta \Gamma$ along with  $\sin 2 \beta$ which characterizes $CP$ violation in the interference of mixing and decay can be masked by the presence of the decoherence parameter. Further, using the Belle analysis of ref. \cite{Belle:2007ocp}, a bound on decoherence parameter was obtained which, till date, is the most stringent bound using B meson systems. In this work, we perform the first analysis of decoherence in purely leptonic B decays. We consider the effect of decoherence on a number of $CP$ violating as well as $CP$ conserving observables related to $B \to \ell \ell$ ($\ell = e, \, \mu,\, \tau$) decays.  This includes analyzing well established relation between theoretical and experimental values of $B(B_s \to \mu^+ \mu^-)$, as obtained \cite{DeBruyn:2012wj,DeBruyn:2012wk}, in the presence of decoherence. Further, we also provide the first analysis of a number of observables in semileptonic decays of a  B meson, both time-dependent as well as independent asymmetries within the Kraus operator formalism which can be used to determine the decoherence parameter. We also revisit our analysis of $B-\bar{B}$ mixing and semileptonic decays of correlated B mesons and provide a theoretical description in the presence of $CP$ violation in mixing. Finally, we also revisit the determination of the decoherence parameter using nonleptonic B decays. Apart from providing theoretical expressions of the time-dependent asymmetries in the presence of $CP$ violation in mixing, we also consider time-independent observables.

Given the fact that many of these observables can be measured very precisely using the wealth of data collected by LHCb and Belle II, our formalism can be utilized to obtain bounds on the decoherence parameter using multiple decay channels. Many of these channels are not only measured with unprecedented precision but also are theoretically clean observables. Thus experiments which were mainly designed to probe new physics can also be utilized to probe physics at the finer length scales. Further, it can also provide an alternative set-up for such studies which at present, are mainly performed in the neutrino sector \cite{Lisi:2000zt,Fogli:2003th,Barenboim:2004wu,Morgan:2005gb,Anchordoqui:2005gj,Capolupo:2018hrp,Buoninfante:2020iyr,DeRomeri:2023dht,IceCube:2023gzt}.

The structure of this study is outlined as follows. In Sec. \ref{open}, we present the temporal evolution of both individual and correlated B mesons within the context of an open quantum system, employing the Kraus operator formalism. Sec. \ref{mmbar} delves into the examination of the impact of decoherence on meson-antimeson mixing. The influence of decoherence on observables associated with semileptonic decays of B mesons is scrutinized in Sec. \ref{sl}. Sec. \ref{lep} is dedicated to an analysis of decoherence effects on observables connected to $B\to \ell \ell$ decays, while Sec. \ref{nl} conducts an examination of non-leptonic B decays. The findings and overall conclusions are summarized in Sec. \ref{concl}. 

\section{Open time evolution of $B$ mesons}
\label{open}

The influence of an ambient environment on the system of interest, such as the $B$ mesons, can be effectively described using the formalism of Open Quantum Systems. However, dealing with $B$ mesons presents additional complexities as they are unstable and decay over time. This introduces a modification in the Hilbert space of the decaying system. Consider pairs of correlated $B^0_d \bar{B^0_d}$ mesons originating from a $\Upsilon(4S)$ resonance, for example. Initially, each pair exists in a joint two-particle Hilbert space. However, due to the inherent decaying nature of the mesons, the Hilbert space evolves to become a combination of two-particle, one-particle, and zero-particle Hilbert spaces. Incorporating this concept into the framework of open quantum systems is essential in our approach to consider the effect of the environment on $B$ meson decays.

\subsection{Time evolution of single $B$ mesons}
\label{singleB}
For single mesons, the Hilbert space can be represented as $\mathcal{H}_{B^0} \oplus \mathcal{H}_0$, where $\mathcal{H}_{B^0}$ is a one-particle Hilbert space, and $\mathcal{H}_0$ is the space spanned by $\ket{0}$, representing the vacuum state and indicating the absence of any particle. The decaying meson system then exists in a Hilbert space spanned by the orthonormal vectors $\ket{B^0}$, $\ket{\bar{B^0}}$, and $\ket{0}$, representing the different possible states of the system during its evolution.:
\be
\ket{B^0} = \left(\begin{array}{c} 1 \\ 0  \\ 0\end{array}\right); \qquad
\ket{\bar{B^0}} =  \left(\begin{array}{c} 0 \\ 1  \\ 0\end{array}\right); \qquad
\ket{0} =  \left(\begin{array}{c} 0 \\ 0 \\ 1\end{array}\right).
\ee
Here $B^0$ stands for $B^0_d/B^0_s$ mesons.

 The meson state is represented by the
density matrix $\rho$. Since, decay as well as decoherence, due to the effect of the environment, is an irreversible
process, the time evolution of $\rho$ can be accounted for by constructing a one-parameter semigroup $\Xi_t$ such that
\be
\rho(t) = \Xi_t \, \rho(0).
\ee
Here $\rho(0)$ denotes the initial state while $\Xi_t$ is a linear superoperator generating the semigroup evolution and is
characterized by 
\be
\mathrm{Tr}\left[\rho(0)\right] = \mathrm{Tr}\left[\Xi_t \,\rho(0) \right] = 1,
\ee
i.e., trace is preserved and 
\be
\Xi_{t_1 + t_2} = \Xi_{t_1}\,\Xi_{t_2},~~ \mathrm{for\, all} \, t_1, t_2 \geq 0.
\ee
From these conditions as well as the condition of complete positivity, the structure of $\Xi_t$ can be found leading to 
$\rho(t)$. The time evolution of the system, consistent with the Geiger-Nutall law, is given by a 
family of completely positive trace-preserving maps forming a one-parameter dynamical
semigroup. Complete positivity implies that the time evolution of a state of the system 
can be written in the operator-sum representation
\be
\rho(t)=\sum_{i=0}^{5} E_i(t) \rho(0) E^{\dagger}_i(t), \label{kraus}
\ee
where $E_i(t)$ are the Kraus operators which  are obtained by comparing 
eq. (\ref{kraus}) with $\rho(t) = \Xi_t \, \rho(0)$. 

The form of the Kraus operators are obtained to be
\bea 
E_0 (t) &=& \ket{0} \bra{0}\,,\\
E_1 (t) &=& \frac{1}{2} \left[ e^{-\left(2 i m_L + \Gamma_L + \lambda \right)t/2} + e^{-\left(2 i m_H + \Gamma_H + \lambda \right)t/2} \right] \left(\ket{B^0} \bra{B^0}  + \ket{\bar{B^0}} \bra{\bar{B^0}} \right)\nonumber \\
&+& \frac{1}{2} \left[ e^{-\left(2 i m_L + \Gamma_L + \lambda \right)t/2} - e^{-\left(2 i m_H + \Gamma_H + \lambda \right)t/2} \right] \left(\frac{p}{q}\ket{B^0} \bra{\bar{B^0}}  
+ \frac{q}{p}\ket{\bar{B^0}} \bra{B^0} \right)\,, \\
E_2 (t) &=& \sqrt{\frac{{\rm Re}[(p-q)/(p+q)]}{|p|^2-|q|^2} \left(1-e^{-\Gamma_L\,t}-\left(|p|^2-|q|^2\right)^2 \frac{\left|1-e^{-(\Gamma+\lambda-i\Delta m)t}\right|^2}{(1-e^{-\Gamma_H\,t})}\right)}\nonumber\\
&& \times \left(\frac{p+q}{2p}\ket{0} \bra{B^0}  + \frac{p+q}{2q}\ket{0} \bra{\bar{B^0}} \right)\,,\\
E_3 (t) &=& \sqrt{\frac{{\rm Re}[(p-q)/(p+q)]}{(|p|^2-|q|^2)(1-e^{-\Gamma_H\,t})}}\Big(\frac{1-e^{-\Gamma_H\,t} + \left(1-e^{-(\Gamma+\lambda-i\Delta m)t}\right)\left(|p|^2-|q|^2\right)}{2p} (p+q)\ket{0} \bra{B^0}\nonumber\\
&& -\frac{1-e^{-\Gamma_H\,t} - \left(1-e^{-(\Gamma+\lambda-i\Delta m)t}\right)\left(|p|^2-|q|^2\right)}{2q} (p+q)\ket{0} \bra{\bar{B^0}}\Big),\\
E_4 (t) &=& \frac{1}{2} e^{-\Gamma_L t /2 }\sqrt{1-e^{-\lambda t}}\left(\ket{B^0} \bra{B^0}  + \ket{\bar{B^0}} \bra{\bar{B^0}} +\frac{p}{q} \ket{B^0} \bra{\bar{B^0}}  + \frac{q}{p}\ket{\bar{B^0}} \bra{B^0} \right)\,,\\
E_5 (t) &=& \frac{1}{2} e^{-\Gamma_H t /2 }\sqrt{1-e^{-\lambda t}}\left(\ket{B^0} \bra{B^0}  + \ket{\bar{B^0}} \bra{\bar{B^0}} - \frac{p}{q}\ket{B^0} \bra{\bar{B^0}} - \frac{q}{p}\ket{\bar{B^0}} \bra{B^0} \right)\,.
\eea
The initial states are represented as
\be 
\rho_{B^0}(0)\equiv \ket{B^0}\bra{B^0}=
\left(\begin{array}{ccc} 
1 & 0 & 0   \\ 0 & 0 & 0   \\ 0 &0 & 0
\end{array}\right); \qquad
\rho_{\bar{B^0}}(0)\equiv \ket{\bar{B^0}}\bra{\bar{B^0}}=
\left(\begin{array}{ccc} 
0 & 0 & 0   \\ 0 &1 & 0   \\ 0 &0 & 0 
\end{array}\right).
\ee
Using the Kraus operators, we obtain
\bea
\rho_{B^0}(t)&=& \frac{1}{2} e^{-  \Gamma t}
\left(\begin{array}{ccc} 
 a_{ch} + e^{-  \lambda t} a_c &  (\frac{q}{p})^*(- a_{sh} -i  e^{-  \lambda t} a_s) & 0  \\ 
(\frac{q}{p})(- a_{sh} + i e^{-  \lambda t} a_s) & |\frac{q}{p}|^2 a_{ch} - e^{-  \lambda t} a_c  & 0 \\ 
0 &0   & \rho_{33}(t)
\end{array}\right),
\nonumber\\
\rho_{\bar{B^0}}(t)&=& \frac{1}{2} e^{-  \Gamma t}
\left(\begin{array}{ccc} 
|\frac{p}{q}|^2 (a_{ch} - e^{-  \lambda t} a_c) & (\frac{p}{q})(- a_{sh} +i  e^{-  \lambda t} a_s) & 0  \\ 
(\frac{p}{q})^* (- a_{sh} - i e^{- \lambda t} a_s) & a_{ch} + e^{-  \lambda t} a_c  & 0 \\ 
0 &0   & \rho'_{33}(t)
\end{array}\right),
\label{dm-bbbar}
\eea
where $a_{ch}=\cosh \left(\frac{ \Delta \Gamma t}{2}\right)$, $a_{sh}=\sinh \left(\frac{ \Delta \Gamma t}{2}\right)$,
$a_c=\cos\left(\Delta m t\right)$ and $a_s=\sin\left(\Delta m t\right)$. $\Gamma=(\Gamma_L + \Gamma_H)/2$, $\Delta \Gamma = 
\Gamma_L -\Gamma_H$,  $\Gamma_L$ and $\Gamma_H$ are the decay width of $B^0_L = p B^0 + q \bar{B^0}$ and $B^0_H = p B^0 - q\bar{B^0}$, respectively (light and heavy 
states of B mesons) with $p$ and $q$ being the complex coefficients that satisfying the condition $|p|^2+|q|^2 = 1$. Further, $\Delta m = m_H - m_L$, $m_L$ and $m_H$ are the masses of $B^0_L$ and $B^0_H$ states, respectively. 
Here $\lambda$ is a decoherence parameter, representing the interaction between the one-particle system and its environment.
$\rho_{33}(t)$ and $ \rho'_{33}(t)$ are some functions of $B$ physics parameters defined above. However, they do not contribute 
to the present analysis. The Kraus operators, initially developed for the $K$ meson system in ref. \cite{Caban:2007je}, have been utilized in our investigation to explore the $B$ meson system.

\subsubsection{Theoretical limit on the decoherence parameter $\lambda$}

For the time evolution of the density operator to be completely positive, the following inequality must hold for any  $t \geq 0$ \cite{Caban:2005ue}:
\begin{equation}
\delta^2_L \left( 1-2e^{- (\Gamma +\lambda)t } \cos(\Delta m t)+e^{-2(\Gamma+\lambda)t}\right) \leq \left(1-e^{-\Gamma_H t}\right) \left(1-e^{-\Gamma_L t}\right)\,.
\end{equation}
Here $\delta_L=|p|^2-|q|^2$. Given that the parameters $\Gamma_L$, $\Gamma_H$, $\Delta m$, and $\delta_L$ are known, the above inequality can be utilized to obtain an upper bound on the decoherence parameter $\lambda$.  This bound serves as a theoretical upper limit, and any experimentally determined limit on $\lambda$ should not exceed this bound. Hence, this inequality also imposes constraints on the validity of the model under consideration. 

For $t \geq 0$ and $\lambda \geq 0$, the above inequality can be written in the following form \cite{Caban:2005ue}
\begin{equation}
    0 \leq \lambda \leq \lambda_{\rm max} \simeq
    \frac{1}{\delta_L}\sqrt{\Gamma_L \Gamma_H -\delta^2_L (\Delta m)^2} - \Gamma \,.
    \label{cond}
\end{equation}

For the neutral $B_d$ meson system, $\Delta m_{B_d} = (0.5065 \pm 0.0019)\, {\rm ps^{-1}}$, $\tau_{B_d}= 1/\Gamma_{B_d}=(1.519 \pm 0.004)\, {\rm ps}$ and $\delta_L = (-0.5 \pm 0.4) \times 10^{-3}$. These values are taken from Particle Data Group (PDG) \cite{pdg}.  As for this system $\Delta \Gamma = \Gamma_L -\Gamma_H \approx 0$, we assume $\Gamma_L\approx\Gamma_H\approx \Gamma $. Substituting these values in eq. \eqref{cond}, we get $\lambda_{\rm max}= 7.9 \times 10^{14}\, {\rm s^{-1}}$ at 95\% C.L.  The current experimental upper
 limit on $\lambda$ is $3\times 10^{10}$ $\rm s^{-1}$ at 95\% C.L. \cite{Alok:2015iua} which is well within the allowed limit. 

For the neutral $B_s$ meson system, $\Delta m_{B_s} = (17.765 \pm 0.006)\, {\rm ps^{-1}}$, $\Gamma_{B_s}=(65.73 \pm 0.23) \times 10^{10}\, {\rm s^{-1}}$, $\tau_{B_{sL}}=1/\Gamma_{B_{sL}}=(1.431 \pm 0.007)\, {\rm ps}$, $\tau_{B_{sH}}=1/\Gamma_{B_{sH}}=(1.624 \pm 0.009)\, {\rm ps}$ and $\delta_L = (-0.15 \pm 0.70) \times 10^{-3}$. By substituting these values, taken from PDG \cite{pdg}, into eq. \eqref{cond}, we derive $\lambda_{\rm max}= 3.64  \times 10^{16}\, {\rm s^{-1}}$ at 95\% C.L.  Currently, there are no experimental limits available for the decoherence parameter in the $B_s$ meson system. However, like the $B_d$ meson system it is expected that the experimental limit on $\lambda$ in $B_s$ system should also fall within the acceptable range.

\subsection{Time evolution of correlated $B$ mesons}
\label{corrB}
 The entangled $B^0-\bar{B}^0$ state can be written as
\be
\ket{\psi(0)} =\frac{1}{\sqrt{2}} \left(\ket{B^0 \bar{B^0}}-\ket{\bar{B^0} B^0}\right)\,.
\label{ini}
\ee
The time evolution of the above state is described by the following density matrix 
\cite{Huet:1994kr,Benatti:1999cq,Benatti:2001tv}:
\be
\rho(t_1,t_2)=\frac{1}{2} \Big(\rho_1(t_1)\otimes \rho_2(t_2) + \rho_2(t_1)\otimes \rho_1(t_2)
- \rho_3(t_1)\otimes \rho_4(t_2)-\rho_4(t_1)\otimes \rho_3(t_2)\Big)\,,
\label{dmcorr}
\ee
where $\rho_1(t)=\rho_{B^0}(t)$, $\rho_2(t)=\rho_{\bar{B^0}}(t)$ which are given in eq.~(\ref{dm-bbbar}),
while $\rho_{3/4}(t) = \sum_i E_i (t)\rho_{3/4}(0) E_i^{\dagger}(t)$, where 
\be 
\rho_{3}(0)\equiv \ket{B^0}
\bra{\bar{B^0}}=
\left(\begin{array}{ccc} 
0 & 1 & 0   \\ 0 & 0 & 0   \\ 0 &0 & 0
\end{array}\right); \qquad
\rho_{4}(0)\equiv  \ket{\bar{B^0}}
\bra{B^0}=
\left(\begin{array}{ccc} 
0 & 0 & 0   \\ 1 &0 & 0   \\ 0 &0 & 0 
\end{array}\right).
\ee
Thus $\rho_{3/4}(t)$ are given by
\bea
\rho_3(t)&=& \frac{1}{2} e^{-  \Gamma t}
\left(\begin{array}{ccc} 
(\frac{p}{q})^* (-a_{sh} -i e^{-  \lambda t} a_s) &  a_{ch} + e^{-  \lambda t} a_c & 0  \\ 
(\frac{p}{q})^* (\frac{q}{p}) (a_{ch} - e^{-  \lambda t} a_c) & (\frac{q}{p}) (-a_{sh} +i e^{-  \lambda t} a_s) & 0 \\ 
0 &0   & \tilde{\rho}_{33}(t)
\end{array}\right),
\nonumber\\
\rho_4(t)&=& \frac{1}{2} e^{-  \Gamma t}
\left(\begin{array}{ccc} 
(\frac{p}{q}) (-a_{sh} +i e^{-  \lambda t} a_s) &  (\frac{p}{q}) (\frac{q}{p})^* (a_{ch} - e^{-  \lambda t} a_c) & 0  \\ 
a_{ch} + e^{-  \lambda t} a_c & (\frac{q}{p})^* (-a_{sh} -i e^{-  \lambda t} a_s)   & 0 \\ 
0 &0   & \tilde{\rho}'_{33}(t)
\end{array}\right).
\eea
Here the parameters are the same as in eq.~(\ref{dm-bbbar}).  $ \tilde{\rho}_{33}(t)$ and $\tilde{\rho}'_{33}(t)$ are functions of $B$ physics parameters defined in the previous subsection and are irrelevant for the present analysis. In the following sections, we discuss various $B$ physics observables that can be affected due to decoherence.

\section{Meson anti-meson mixing}
\label{mmbar}
In the presence of decoherence, the survival (oscillation) probability 
of initial $B^0$ meson to decay as $B^0 (\bar{B}^0)$ at a proper decay 
time $t$ is given by
\bea
P_{+} (t,\lambda) &=& \frac{e^{-\Gamma\, t}}{2}\left[\cosh(\Delta \Gamma\, t/2) 
+ e^{-\lambda\, t} \cos (\Delta m\, t)\right]\, ,\nonumber\\
P_{-} (t,\lambda) &=&\frac{e^{-\Gamma\, t}}{2} \left|\frac{q}{p}\right|^2 \left[\cosh(\Delta \Gamma\, t/2) 
- e^{-\lambda\, t} \cos (\Delta m\, t)\right]\,,
\label{spm}
\eea
respectively. The positive sign applies when the $B^0$ meson decays with the same flavor as its 
production and the negative sign when the particle decays with an opposite flavor to 
its production. We see that the survival (oscillation) probability of $B^0$ is 
$\lambda$ dependent! The time-dependent mixing asymmetry, used to determine $\Delta m$ and $\Delta \Gamma$, 
is defined as
\be 
A_{\rm mix} (t,\lambda)\equiv \frac{P_{+} (t,\lambda)-P_{-} (t,\lambda)}{P_{+} (t,\lambda)+P_{-} (t,\lambda)} \,. 
\label{amix-asy}
\ee
This asymmetry can be observed using a flavor tagging decay, like $B^0 \to D^- l^+ \nu$.
Using eqs.~(\ref{spm}), $A_{\rm mix} (t,\lambda)$ is given by
\be 
A_{\rm mix} (t,\lambda)=
\frac{\left(1 - \left|\frac{q}{p}\right|^2\right)\cosh (\Delta \Gamma\, t/2) + \left(1+ \left|\frac{q}{p}\right|^2\right)e^{-\lambda\, t} \cos (\Delta m\, t)}
{\left(1 + \left|\frac{q}{p}\right|^2\right)\cosh (\Delta \Gamma\, t/2) + \left(1 - \left|\frac{q}{p}\right|^2\right)e^{-\lambda\, t} \cos (\Delta m\, t) } \,. 
\label{amix0}
\ee
The above equation can also be written as
\be 
A_{\rm mix} (t,\lambda)=
\frac{e^{-\lambda\, t} \cos (\Delta m\, t) + \delta_B\, \cosh (\Delta \Gamma\, t/2)}
{\cosh (\Delta \Gamma\, t/2) + \delta_B \, e^{-\lambda\, t} \cos (\Delta m\, t) } \,,
\label{amix}
\ee
where $\delta_B$ is defined as 
\be
\delta_B \equiv \frac{1 - \left|\frac{q}{p}\right|^2}{1 + \left|\frac{q}{p}\right|^2}\,.
\ee
In the limit of neglecting the $CP$ violation in mixing,  i.e., $|q/p|=1$, we get
\be 
A_{\rm mix} (t,\lambda)=
\frac{ \cos (\Delta m\, t) }{\cosh (\Delta \Gamma\, t/2) } e^{-\lambda\, t} \,.
\label{amix1}
\ee
Thus we see that the decoherence parameter appears in the expression of the time-dependent mixing asymmetry $A_{\rm mix} (t)$. 
The true value of $\Delta m$, along with $\Delta \Gamma$, can be determined by 
a three parameter ($\Delta m,\, \Delta \Gamma,\,\lambda$) fit to  $A_{\rm mix} (t,\lambda)$ given in eq.~(\ref{amix0}) or eq.~(\ref{amix}). 

The $B^0_d-\bar{B^0_d}$ mixing was first observed in 1987 through time-integrated measurements by the UA1 \cite{UA1:1986fuh} and ARGUS \cite{UA1:1986fuh} collaborations. These measurements typically relied on the enumeration of lepton pairs with either the same or opposite signs originating from the semileptonic decay of the generated $b\bar{b}$ pairs. Subsequently, numerous other experiments have also contributed to the measurements of $B^0_d-\bar{B^0_d}$ mixing, both through time-dependent and time-integrated analyses. The experimental determination of $\Delta m_d$ can be achieved through various approaches. The LHCb, CDF, and D0 experiments employ a methodology that involves measuring the decay rates of a particle initially identified as a pure $B^0_d$ at time $t=0$ which subsequently decays either as a $B^0_d$ or  $\bar{B^0_d}$ over the course of its proper decay time. Conversely, the Belle and BaBar experiments adopt a different approach to determine $\Delta m_d$. They do so by measuring the time-dependent probabilities, denoted as $P_+(t)$ for unoscillated $B^0_d \bar{B^0_d}$ events and $P_-(t)$ for oscillated $B^0_d B^0_d$/$\bar{B^0_d} \bar{B^0_d}$ events, involving two neutral $B_d$ mesons that originate from an entangled state during the decay of the $\Upsilon(4S)$ resonance. The expressions for $P_\pm (t)$,  in the presence of decoherence, remain consistent with those provided in eq.~\eqref{spm}, albeit with substitution of the proper time $t$ with the proper decay-time difference $\Delta t$ between the decays of the two neutral $B_d$ mesons.

The current world average for $\Delta m_d$, as cited in the PDG, stands at $(0.5065\pm 0.0019),\rm ps^{-1}$ \cite{pdg}. This value represents the amalgamation of measurements of $\Delta m_d$ derived from a spectrum of experiments, including OPAL \cite{opal}, ALEPH \cite{aleph}, DELPHI \cite{delphi}, L3 \cite{l3}, CDF \cite{cdf}, BaBar \cite{babar}, Belle \cite{belle}, D0 \cite{do}, and LHCb \cite{LHCb:2011vae,LHCb:2012mhu,LHCb:2013fep}. Consequently, it becomes apparent that the process of determining $\Delta m_d$ in experiments like LHCb, CDF, D0, Belle, and BaBar can be affected by the presence of the decoherence parameter $\lambda$. Therefore, reevaluating the determination of $\Delta m_d$ in these experiments while considering decoherence can provide constraints on the $\lambda$ parameter. On the other hand, the LEP experiments gauge $\Delta m_d$ through time-independent measurements.  In the following section, we will demonstrate that these measurements are likewise affected by the presence of decoherence.

The first observation of $B_s^0 - \bar{B_s^0}$ mixing was reported in 2006 by the CDF collaboration \cite{CDF:2006imy}, which was 19 years after the initial detection of $B_d^0 - \bar{B_d^0}$ mixing. This delay is primarily attributed to the significant discrepancy in oscillation frequencies between the $B_s^0 - \bar{B_s^0}$ and $B_d^0 - \bar{B_d^0}$ systems. The former has an oscillation frequency 35 times greater than the latter, creating a substantial challenge for achieving precise decay time resolution in detectors. LHCb subsequently detected oscillations in $B_s^0 - \bar{B_s^0}$ by utilizing data spanning from 1 to 4.9 $\rm fb^{-1}$ collected at the LHC up until 2016 \cite{LHCb:2011vae,LHCb:2013lrq,LHCb:2013fep,LHCb:2014iah}. More recently, measurements derived from the complete LHC Run 2 dataset have been released by both the CMS \cite{CMS:2020efq} and LHCb \cite{LHCb:2020qag,LHCb:2021moh} collaborations. The current world average of $\Delta m_s$ stands at $17.765 \pm 0.006$ \cite{pdg}, with the most recent LHCb result contributing significantly to this remarkable level of precision. The process for determining $\Delta m_s$ at both CDF and LHC is essentially the same as that for $\Delta m_d$ with the addition of $\Delta \Gamma_s$ in the fitting process. Consequently, just like $\Delta m_d$, the methods used to determine $\Delta m_s$ are also influenced by the presence of decoherence effects. Therefore, a reevaluation of time-dependent data for $B_s^0 - \bar{B_s^0}$ oscillations not only allows for the determination of the {\it true} values of $\Delta m_s$ and $\Delta \Gamma_s$, but also provides an opportunity to establish constraints on the decoherence parameter. Given the exceptional precision in determining mixing parameters $\Delta m_d$ and $\Delta m_s$ owing to the abundant data available, a reanalysis of this data within the current framework holds the potential for enhancing the existing constraints on $\lambda$. The precision is expected to be further improved at Belle II and HL-LHC, assuming that the flavor tagging performance can be maintained or enhanced \cite{LHCb:2018roe,Belle-II:2018jsg}.

\section{Semileptonic decays of $B$ mesons}
\label{sl}
In the formalism of density matrices, any physical observable of the neutral
$B$-meson system is described by a suitable hermitian operator $\mathcal{O}$.
Its evolution in time can be obtained by taking its trace with the density matrix $\rho(t)$.
Of particular interest are those observables $\mathcal{O}_f$ that are 
associated with the decay of a $B$-meson into final states `$f$'. In the 
$\ket{B^0}$, $\ket{\bar{B^0}}$, and $\ket{0}$ basis, $\mathcal{O}_f$ is represented by
\be 
\mathcal{O}_f = \left(\begin{array}{ccc} 
|A(B^0 \to f)|^2 & {A(B^0 \to f)}^* A(\bar{B^0} \to f) & 0  \\ 
  A(B^0 \to f) {A(\bar{B^0} \to f)}^*&  |A(\bar{B^0} \to f)|^2& 0 \\ 
0 &0   & 0
\end{array}\right).
\ee
Here the entries are written in terms of the two independent 
decay amplitudes $A(B^0 \to f)$ and $A(\bar{B^0} \to f)$. Probability rate, $P_f(B^0/ \bar{B^0};t)$, that an initial $\rho_{B^0/ \bar{B^0}}$ 
state decays at time $t$ into a given state $f$, described by the operator $\mathcal{O}_f$, is then given by
 ${\rm Tr}\,[ \mathcal{O}_f \, \rho(t)]$.

\subsection{Semileptonic decays of single $B^0$ mesons}
The temporal progression of a single B-meson in the presence of an external environment is elaborated upon in Sec. \ref{singleB}. With this formalism at our disposal, we can now pinpoint certain observables whose measurement can facilitate the determination of the decoherence parameter $\lambda$. In this section, we delve into a discussion of these specific observables. Let us consider the decays of $B^0$ mesons in semileptonic states $h\, l\, \nu$, where $h$ stands for any allowed 
charged hadronic state. Under the assumption of $CPT$ conservation and no violation of $\Delta B = \Delta Q$ rule, the amplitudes for $B^0/\bar{B^0}$ into $h^- l^+ \nu$ can be written as 
\be 
A \left(B^0 \to h^- l^+ \nu \right)=M_h\,, \qquad A \left(\bar{B^0} \to h^- l^+ \nu \right)=0\,,
\ee
whereas the amplitudes for $B^0/\bar{B^0}$ into $h^+ l^- \bar{\nu}$ can be written as 
\be 
A\left(B^0 \to h^+ l^- \bar{\nu} \right)=0\,, \qquad A\left(\bar{B^0} \to h^+ l^- \bar{\nu} \right)=M^*_h\,.
\ee
Thus the operators for these decay modes can be written as
\be 
\mathcal{O}_{h^-} = \left(\begin{array}{ccc} 
|M_h|^2 & 0 & 0  \\ 0 &0 &0 \\ 0 & 0 & 0
\end{array}\right), \qquad
\mathcal{O}_{h^+} = \left(\begin{array}{ccc} 
0 & 0  & 0  \\ 0 &|M^*_h|^2 &0 \\ 0 & 0 & 0
\end{array}\right).
\label{opsemilep}
\ee

Therefore the probability rate that an initial state $B^0$ decays into a final 
state $h^- l^+ \nu / h^+ l^- \bar{\nu}$ is given by
\bea
P_{h^-}(B^0;t) &=& \frac{|M_h|^2}{2} e^{- \Gamma t} \left[\cosh \left(\frac{t \Delta \Gamma}{2}\right) + e^{- \lambda t}\cos\left(\Delta m t\right)\right], \nonumber\\
P_{h^+}(B^0;t) &=& \frac{|M^*_h|^2}{2} e^{- \Gamma t}\left|\frac{q}{p}\right|^2 \left[\cosh \left(\frac{t \Delta \Gamma}{2}\right) - e^{- \lambda t}\cos\left(\Delta m t\right)\right].
\label{pbos}
\eea
If the initial state is $\bar{B^0}$ then 
\bea
P_{h^-}(\bar{B^0};t) &=& \frac{|M_h|^2}{2} e^{-\Gamma t} \left|\frac{p}{q}\right|^2\left[\cosh \left(\frac{t \Delta \Gamma}{2}\right) - e^{-\lambda t}\cos\left(\Delta m t\right)\right], \nonumber\\
P_{h^+}(\bar{B^0};t) &=& \frac{|M^*_h|^2}{2} e^{- \Gamma t} \left[\cosh \left(\frac{t \Delta \Gamma}{2}\right) + e^{- \lambda t}\cos\left(\Delta m t\right)\right].
\label{pbbaros}
\eea

We can define several $\lambda$ dependent asymmetries using these semileptonic probability rates. However, for determining $\lambda$, only those asymmetries are important which do not vanish in the limit of neglecting $CP$ violation in mixing or $\Delta \Gamma$. Otherwise, the dependence on $\lambda$ will always be suppressed by these quantities making its determination difficult. 
For example, we can see that the following semileptonic ``wrong-charge" asymmetry 
\begin{equation}
\frac{P_{h^-}(\bar{B^0};t) - P_{h^+}(B^0;t)}{P_{h^-}(\bar{B^0};t) + P_{h^+}(B^0;t)}=\frac{1-\left|\frac{q}{p}\right|^4}{1+\left|\frac{q}{p}\right|^4}\,,
\end{equation}
is independent of the decoherence parameter $\lambda$. This asymmetry is used to determine the $CP$ violation in mixing, i.e., it has been subject to measurement through both time-integrated analyses conducted at CLEO \cite{CLEO:2000sdf}, BaBar \cite{BaBar:2014bbb}, CDF \cite{CDF:1996kan},DØ \cite{D0:2013ohp} and LHCb \cite{LHCb:2016ssr}, as well as time-dependent analyses carried out at LEP \cite{ALEPH:2000las}, BaBar \cite{BaBar:2013xng} and Belle \cite{Belle:2005cou}. The average of measurements of the above wrong-charge asymmetry leads to $|q/p|_d=1.0010 \pm 0.0008$ and $|q/p|_s=1.0003 \pm 0.0014$ \cite{pdg}. Here ``$d$" and ``$s$" subscripts are associated with the $B_d^0$ and $B_s^0$ systems, respectively. Therefore, these measurements will not be modulated in the presence of decoherence. 

We can now  define the following $\lambda$ dependent asymmetries
\bea 
A^1_{\lambda} (t) &\equiv& \frac{P_{h^-}(B^0;t) - P_{h^-}(\bar{B^0};t)}{P_{h^-}(B^0;t) + P_{h^-}(\bar{B^0};t)},\\
A^2_{\lambda} (t) &\equiv& \frac{P_{h^+}(B^0;t) - P_{h^+}(\bar{B^0};t)}{P_{h^+}(B^0;t) + P_{h^+}(\bar{B^0};t)}.
\eea
Using the expression of probability rates given above, we obtain
\bea 
A^1_{\lambda} (t) &=& \frac{e^{-\lambda\, t} \cos (\Delta m\, t) - \delta_B\, \cosh (\Delta \Gamma\, t/2)}{\cosh (\Delta \Gamma\, t/2) - \delta_B\, e^{-\lambda\, t} \cos (\Delta m\, t)},\\
A^2_{\lambda} (t) &=& - \frac{e^{-\lambda\, t} \cos (\Delta m\, t) + \delta_B\, \cosh (\Delta \Gamma\, t/2) }{\cosh (\Delta \Gamma\, t/2) + \delta_B\, e^{-\lambda\, t} \cos (\Delta m\, t)}.
\eea
Thus decoherence parameter $\lambda$ can be determined  from the measurement of above time dependent asymmetries.
Neglecting $CP$ violation in mixing, above asymmetries takes a simpler form,
\bea 
A^1_{\lambda} (t) &=& e^{- \lambda t } \frac{\cos\left(\Delta m t\right)}{\cosh \left(\frac{ \Delta \Gamma t}{2}\right)},\\
A^2_{\lambda} (t) &=& - e^{- \lambda t } \frac{\cos\left(\Delta m t\right)}{\cosh \left(\frac{ \Delta \Gamma t}{2}\right)}.
\eea
Hence, the dependence of these asymmetries on the decoherence parameter $\lambda$ remains significant even as the limit of $|q/p| \rightarrow 1$ is approached.
The above asymmetries can be translated into integrated asymmetries as
\bea 
A^1_{\lambda}  &=& \frac{P_{h^-}(B^0) - P_{h^-}(\bar{B^0})}{P_{h^-}(B^0) + P_{h^-}(\bar{B^0})},\\
A^2_{\lambda}  &=& \frac{P_{h^+}(B^0) - P_{h^+}(\bar{B^0})}{P_{h^+}(B^0;t) + P_{h^+}(\bar{B^0})},
\eea
where $P_f(B)=(1/\Gamma)\int_0^\infty P_f(B;t)\,dt$. Using eqs. (\ref{pbos}) and (\ref{pbbaros}), we get
\bea
A^1_{\lambda}  &=& \frac{(1+\lambda')((1+\lambda')^2+x^2)^{-1}-\delta_B\,(1-y^2)^{-1}}{(1-y^2)^{-1} -\delta_B\,(1+\lambda')((1+\lambda')^2+x^2)^{-1}},\\
A^2_{\lambda}  &=& - \frac{(1+\lambda')((1+\lambda')^2+x^2)^{-1}+\delta_B\,(1-y^2)^{-1}}{(1-y^2)^{-1} +\delta_B\,(1+\lambda')((1+\lambda')^2+x^2)^{-1}},
\eea
where  $x=\Delta m/\Gamma$, $y=\Delta \Gamma/2\Gamma$ and $\lambda'=\lambda/\Gamma$. Thus these integrated symmetries can also be utilized to measure the decoherence parameter. 

\subsection{Semileptonic decays of correlated $B^0$ mesons}
The double decay rate, $G(f,t_1;g,t_2)$, 
that the left-moving meson decays at proper time $t_1$ into a final state $f$,
while the right-moving meson decays at proper time $t_2$ into the final state $g$,  
is then given by ${\rm Tr}\,[ (\mathcal{O}_f \otimes \mathcal{O}_g ) \, \rho(t_1,t_2)]$. 
From this, a very useful quantity called the single time distribution, $\Gamma(f,g;t)$, can be defined as 
 $\Gamma(f,g;t)=\int_0^{\infty} d\tau\, G(f,\tau+t;g,\tau)$, where $t=t_1-t_2$ is taken to be positive.

We now consider the following time-dependent observables
\bea
 R(t) &\equiv& \frac{\Gamma(h^+,h^+;t)+\Gamma(h^-,h^-;t)}{\Gamma(h^+,h^-;t)+\Gamma(h^-,h^+;t)}\,,\\
 \chi (t) &\equiv& \frac{\Gamma(h^+,h^+;t)+\Gamma(h^-,h^-;t)}{\Gamma(h^+,h^+;t)+\Gamma(h^-,h^-;t)+\Gamma(h^+,h^-;t)+\Gamma(h^-,h^+;t)}\,.
\eea

Using (\ref{dmcorr}) and (\ref{opsemilep}), we get
\bea
 R(t) &=& \frac{1}{2}\left(\left|\frac{p}{q}\right|^2 + \left|\frac{q}{p}\right|^2\right) \frac{(1+\lambda')\cosh \left(\frac{ \Delta \Gamma t}{2}\right)-e^{-\lambda\,t}\, \cos\left(\Delta m t\right)}{(1+\lambda')\cosh \left(\frac{ \Delta \Gamma t}{2}\right)+e^{-\lambda\,t}\, \cos\left(\Delta m t\right)}\,,
\label{Rt}\\
 \chi (t) &=& \frac{\left(\left|\frac{p}{q}\right|^2 + \left|\frac{q}{p}\right|^2\right)\Big(\cosh \left(\frac{ \Delta \Gamma t}{2}\right)-(1+\lambda')^{-1}e^{-\lambda\,t}\cos\left(\Delta m t\right)\Big)}{\left(2-\left|\frac{p}{q}\right|^2 -\left|\frac{q}{p}\right|^2\right) (1+\lambda')^{-1}e^{-\lambda\,t}\cos\left(\Delta m t\right)+\left(2+\left|\frac{p}{q}\right|^2 +\left|\frac{q}{p}\right|^2\right)\cosh \left(\frac{ \Delta \Gamma t}{2}\right)}\,.
 \label{chit}
\eea
We see that the above observables do not vanish in the limit of neglecting $CP$ violation in mixing and $\Delta \Gamma$ and hence useful for the determination of decoherence parameter $\lambda$. One can also define the following observables
\bea 
A^{{\rm corr},\,1}_{\lambda} (t) &\equiv&\frac{\Gamma(h^+,h^-;t) - \Gamma(h^+,h^+;t)}{\Gamma(h^+,h^-;t)+\Gamma(h^+,h^+;t)},\\
A^{{\rm corr},\,2}_{\lambda} (t) &\equiv& \frac{\Gamma(h^-,h^+;t)-\Gamma(h^-,h^-;t)}{\Gamma(h^-,h^+;t)+\Gamma(h^-,h^-;t)}.
\eea
Using (\ref{dmcorr}) and (\ref{opsemilep}), we get
\bea 
A^{{\rm corr},\,1}_{\lambda} (t) &=& \frac{(1+\lambda')^{-1}\,e^{-\lambda\,t}\cos\left(\Delta m t\right) + \delta_B\,\cosh \left(\frac{ \Delta \Gamma t}{2}\right)}
{\cosh \left(\frac{ \Delta \Gamma t}{2}\right) + \delta_B\,(1+\lambda')^{-1}\,e^{-\lambda\,t}\cos\left(\Delta m t\right) },\\
A^{{\rm corr},\,2}_{\lambda} (t) &=& \frac{(1+\lambda')^{-1}\,e^{-\lambda\,t}\cos\left(\Delta m t\right) -\delta_B\,\cosh \left(\frac{ \Delta \Gamma t}{2}\right)}
{\cosh \left(\frac{ \Delta \Gamma t}{2}\right) - \delta_B\,(1+\lambda')^{-1}\,e^{-\lambda\,t}\cos\left(\Delta m t\right)}.
\eea
In the limit of neglecting $CP$ violation in mixing, the above equation takes the following simple form
\be
A^{{\rm corr},\,1}_{\lambda} (t) = A^{{\rm corr},\,2}_{\lambda} (t)  = \left(\frac{1}{1 + \lambda'}\right) e^{-\lambda t} \frac{\cos\left(\Delta m t\right)}{\cosh \left(\frac{ \Delta \Gamma t}{2}\right)}\,.
\ee

We now turn our attention towards time-independent observables. There are two important observables which can be affected by interaction with the environment.
 One is the ratio of the total same-sign to opposite-sign semileptonic rates
 \be
 R = \frac{\Gamma(h^+,h^+)+\Gamma(h^-,h^-)}{\Gamma(h^+,h^-)+\Gamma(h^-,h^+)}\,,
 \ee
 and the other is the total $B^0 - \bar{B^0}$ mixing probability
 \be
 \chi = \frac{\Gamma(h^+,h^+)+\Gamma(h^-,h^-)}{\Gamma(h^+,h^+)+\Gamma(h^-,h^-)+\Gamma(h^+,h^-)+\Gamma(h^-,h^+)}\,.
 \ee
Time-independent probabilities, $\Gamma(f,g)$, can be obtained by integrating 
the distribution $\Gamma(f,g;t)$ over time. The expressions for $R$ and $\chi$ are obtained to be 
 \bea
 R &=& \frac{1}{2}\left(\left|\frac{p}{q}\right|^2 + \left|\frac{q}{p}\right|^2\right) 
 \frac{(1-y^2)^{-1} - \left((1+\lambda')^2 +x^2\right)^{-1}}{(1-y^2)^{-1} + \left((1+\lambda')^2 +x^2\right)^{-1}}\,,
 \label{R}\\
 \chi &=& \frac{\left(\left|\frac{p}{q}\right|^2 + \left|\frac{q}{p}\right|^2\right) \left( (1-y^2)^{-1} - \left((1+\lambda')^2 +x^2\right)^{-1}\right)}
 {\left(2+\left|\frac{p}{q}\right|^2 + \left|\frac{q}{p}\right|^2\right)(1-y^2)^{-1}+\left(2-\left|\frac{p}{q}\right|^2- \left|\frac{q}{p}\right|^2\right)\left((1+\lambda')^2 +x^2\right)^{-1}}\,.
 \label{chi}
 \eea
 Thus we see that these quantities are $\lambda$ dependent. The LEP experiments utilized the measurements of these time-independent observables to determine the value of  $\Delta m_d$. As these quantities are $\lambda$ dependent,  a reevaluation of the LEP data using the above expressions will enable the imposition of constraints on the possible values of $\lambda$.

However, another well-known asymmetry, the $CP$ asymmetry $A_{\rm CP}$ between $P (\bar{B} \to B)$ and $P (B \to \bar{B})$  which can be measured by the following charge asymmetry of the same-sign
dilepton event rate 
\begin{equation*}
    A_{\rm CP}= \frac{\Gamma(h^+,h^+)-\Gamma(h^-,h^-)}{\Gamma(h^+,h^-)+\Gamma(h^-,h^+)}=\frac{\left|\frac{p}{q}\right|^2 - \left|\frac{q}{p}\right|^2}{\left|\frac{p}{q}\right|^2 + \left|\frac{q}{p}\right|^2}\,,
\end{equation*}
turns out to be independent of $\lambda$.

\section{Purely leptonic decays of $B$ mesons}
\label{lep}
In this section, we study the effects of decoherence on observables related to the purely leptonic decays of $B$  mesons. The quark level transition $b \to (s,d) \ell \ell$ ($\ell=e,\,\mu,\,\tau$) induces purely leptonic decays $B_q^0 \to \ell \ell$. These decays apart from being loop suppressed are also chirally suppressed within the SM. We particularly focus on $B^0_s \to \mu^+ \, \mu^-$ decay. 
The SM prediction for the branching ratio of $B^0_s \to \mu^+ \, \mu^-$ is $(3.47 \pm 0.14)\times 10^{-9}$ \cite{Bobeth:2013uxa,UTfit:2022hsi}. On the experimental side,  $B^0_s \to \mu^+ \, \mu^-$ has been observed with a branching ratio of $(3.45\pm0.29) \times 10^{-9} $ \cite{Ciuchini:2022wbq}. The world average value was revised following an updated measurement of the branching ratio by the CMS collaboration, utilizing the complete Run II dataset \cite{CMS:2022dbz}. Subsequent to this update, the experimental world average for the branching ratio of $B^0_s \to \mu^+  \mu^-$ aligns with the SM prediction.

In Ref. \cite{DeBruyn:2012wj}, it was shown that due to the sizable decay width difference $\Delta \Gamma_s$ of the $B^0_s$ meson, the theoretical
branching ratio of $B^0_s \to \mu^+ \, \mu^-$ needs to be rescaled in order to compare it with the experimentally measured branching ratio. In  \cite{DeBruyn:2012wk}, a new observable called
effective $B^0_s \to \mu^+ \, \mu^-$ lifetime, $\tau_{\mu^+\,\mu^-}$, was introduced. Apart from probing new physics, this observable would facilitate the inclusion of 
$\Delta \Gamma_s$ effect in the comparison between the theoretical and experimental branching ratio of $B^0_s \to \mu^+ \, \mu^-$. It would be
interesting to study these observables in the presence of decoherence. 

 The effective Lagrangian for the process  $B^0_s/\bar{B}^0_s \to \ell^+ \, \ell^-$ ($\ell=e,\,\mu\,, \tau$) can be written as \cite{Chankowski:2004tb}
\begin{equation}
\mathcal{L}_{eff} = B^0_s \bar{\psi}_{\ell}(b+a\gamma_5)\psi_{\ell} + \bar{B}^0_s \bar{\psi}_{\ell}(\bar{b} + \bar{a} \gamma_5)\psi_{\ell}.
\label{leff}
\end{equation}
Under the assumption of $CPT$ conservation,  $\bar{b}=b^*$ and $\bar{a}=-a^*$. Helicity conservation in the decay of $B^0_s$ or $\bar{B}^0_s$ implies that the only final states can be $\ell^+_L \, \ell^-_L$ or $\ell^+_R \, \ell^-_R$, which are $CP$ conjugates. We therefore consider  decays $B^0_s/\bar{B}^0_s \to \ell^+_h \, \ell^-_h$, where $h=L,R$. The operator for these processes is given by 
\be 
\mathcal{O}_{h} = |A_{h}|^2\left(\begin{array}{ccc} 
1 & (\frac{p}{q})\,\lambda_{h} & 0  \\ (\frac{p}{q})^*\lambda_{h}^*  &|\frac{p}{q}|^2 |\lambda_h|^2 &0 \\ 0 & 0 & 0
\end{array}\right),
\label{opbsmumu}
\ee
where the phase invariant quantity $\lambda_h$ is  defined as 
\be 
\lambda_h = \frac{q}{p}\frac{\bar{A_h}\left(\equiv A(\bar{B}^0_s \to \ell^+_h \, \ell^-_h)\right)}{A_h \left(\equiv A(B^0_s \to \ell^+_h \, \ell^-_h)\right)}\,.
\ee

Using eq.~\eqref{leff}, we get 
\be 
\lambda_L = \frac{q}{p} \left( \frac{-a^* + \beta b^*}{a+\beta b} \right) \,, \qquad \lambda_R = \frac{q}{p} \left( - \frac{a^* + \beta b^*}{a-\beta b}\right) \,,
\ee
where $\beta=\sqrt{1-4m^2_{\ell}/m^2_B}$.
Therefore the probability rate that an initial state $B^0_s/\bar{B}^0_s$ decays into final 
state $\ell^+_h \, \ell^-_h$ is given by
\bea
\label{plep1}
\frac{P_{h}(B^0_s;t)}{\frac{1}{2} e^{-  \Gamma t}|A_h|^2} &=&
\left(1+|\lambda_h|^2\right) \cosh \left(\frac{ \Delta \Gamma t}{2}\right) 
+\left(1-|\lambda_h|^2\right)e^{-  \lambda t} \cos\left(\Delta m t\right) 
\nonumber\\
&&- 2{\rm Re}(\lambda_h)\sinh \left(\frac{ \Delta \Gamma t}{2}\right)
-2 {\rm Im}(\lambda_h)e^{-  \lambda t} \sin\left(\Delta m t\right)\,,\\
\label{plep2}
\frac{P_{h}(\bar{B}^0_s;t)}{\frac{1}{2} e^{-  \Gamma t}|A_h|^2 |\frac{p}{q}|^2} &=&
\left(1+|\lambda_h|^2\right) \cosh \left(\frac{ \Delta \Gamma t}{2}\right) 
-\left(1-|\lambda_h|^2\right)e^{-  \lambda t} \cos\left(\Delta m t\right) 
\nonumber\\
&&- 2{\rm Re}(\lambda_h)\sinh \left(\frac{ \Delta \Gamma t}{2}\right)
+2 {\rm Im}(\lambda_h)e^{- \lambda t } \sin\left(\Delta m t\right)\,.
\eea

We can now construct the following $CP$ violating asymmetries using the polarized decay probabilities \cite{Huang:2000tz,Chankowski:2004tb,Alok:2011gv}
\bea
A_{CP}^{LR}(t) & \equiv & 
\frac{P(\bs(t) \to \ell^+_L \ell^-_L) - P(\bsbar(t) \to \ell^+_R \ell^-_R) }
{  P(\bs(t) \to \ell^+_L \ell^-_L) + P(\bsbar(t) \to \ell^+_R \ell^-_R) }\; ,
\nonumber \\
A_{CP}^{RL}(t) & \equiv &  
\frac{P(\bs(t) \to \ell^+_R \ell^-_R) - P(\bsbar(t) \to \ell^+_L \ell^-_L)   }
{ P(\bs(t) \to \ell^+_R \ell^-_R) + P(\bsbar(t) \to \ell^+_L \ell^-_L) }\;.
\eea
The $CP$ asymmetry in the longitudinal polarization fraction $A_{LP}$ \cite{Alok:2008hh} may be written in terms of these two CP asymmetries.
The measurement of either of these CP asymmetries for $B^0_s \to \mu^+ \, \mu^-$ requires the measurement of muon polarization, which will be an impossible task for
the upcoming experiments \cite{Alok:2010zd}. However, these asymmetries can be measured for the $B_s \to \tau^+ \tau^-$ decay mode. The numerator and denominator of  $A_{CP}^{LR}(t)$ is calculated to be

\begin{eqnarray}
N_{CP}^{LR}(t) &=& \frac{1}{2} e^{-  \Gamma t} |A_L|^2\Bigg[\left(|\lambda_L|^2 - |\lambda_R|^{-2}\right)\cosh \left(\frac{ \Delta \Gamma t}{2}\right) 
 - \left(|\lambda_L|^2 - |\lambda_R|^{-2}\right) e^{-  \lambda t} \cos\left(\Delta m t\right) \nonumber\\
&&-2{\rm Re}\left(\lambda_L - \lambda_R^{-1}\right)\sinh \left(\frac{ \Delta \Gamma t}{2}\right)
- 2 {\rm Im} \left(\lambda_L - \lambda_R^{-1}\right)e^{- \lambda t } \sin\left(\Delta m t\right)
\Bigg]\,,\nonumber\\
D_{CP}^{LR}(t) &=& \frac{1}{2} e^{-  \Gamma t} |A_L|^2 \Bigg[ \left\{2+ \left(|\lambda_L|^2 + |\lambda_R|^{-2}\right) \right\}\cosh \left(\frac{ \Delta \Gamma t}{2}\right)  \nonumber\\
&&
+\left\{2- \left(|\lambda_L|^2 + |\lambda_R|^{-2}\right) \right\}e^{-  \lambda t} \cos\left(\Delta m t\right) -2 {\rm Re}\left(\lambda_L + \lambda_R^{-1}\right) \sinh \left(\frac{ \Delta \Gamma t}{2}\right) \nonumber\\
&&
-2 {\rm Im} \left(\lambda_L + \lambda_R^{-1}\right) e^{- \lambda t } \sin\left(\Delta m t\right)
\Bigg]\,.
\end{eqnarray}
Using above equations, the integrated asymmetry $A_{CP}^{LR}$ can be obtained as
\begin{equation}
A_{CP}^{LR} = \frac{\int_0^{\infty}N_{CP}^{LR}(t) dt }{\int_0^{\infty}D_{CP}^{LR}(t) dt}\equiv \frac{N_{CP}^{LR}}{D_{CP}^{LR}}\,,
\label{acp-lr}
\end{equation}
where
\begin{eqnarray}
N_{CP}^{LR} &=& \frac{\left(|\lambda_L|^2 - |\lambda_R|^{-2}\right)}{1-y^2}
-\frac{(1+\lambda')\left(|\lambda_L|^2 - |\lambda_R|^{-2}\right)}{(1+\lambda')^2+x^2}
+\frac{2\,y\,{\rm Re}\left(\lambda_L - \lambda_R^{-1}\right)}{1-y^2}-\frac{2\,x\,{\rm Im} \left(\lambda_L - \lambda_R^{-1}\right)}{(1+\lambda')^2+x^2}\,,\nonumber\\
D_{CP}^{LR} &=& \frac{ \left\{2+ \left(|\lambda_L|^2 + |\lambda_R|^{-2}\right) \right\}}{1-y^2}+\frac{(1+\lambda')\left\{2- \left(|\lambda_L|^2 + |\lambda_R|^{-2}\right) \right\}}{(1+\lambda')^2+x^2}
+\frac{2\,y\, {\rm Re}\left(\lambda_L + \lambda_R^{-1}\right)}{1-y^2}
\nonumber\\
&&
-\frac{2\,x\, {\rm Im} \left(\lambda_L + \lambda_R^{-1}\right)}{(1+\lambda')^2+x^2}\,.
\end{eqnarray}
It should be noted that the above formalism is also valid for $B_d \to \ell^+ \ell^-$ decays. We now consider the following cases:
\begin{enumerate}

\item {\it No decoherence, i.e., $\lambda=0$ and $\Delta \Gamma=0$:} Under these assumptions, eq. \eqref{acp-lr} takes the simpler form
\begin{equation}
A_{CP}^{LR}=\frac{\frac{x^2}{2}\left(|\lambda_L|^2 - |\lambda_R|^{-2}\right)-x\,{\rm Im} \left(\lambda_L - \lambda_R^{-1}\right)}{2+x^2+\frac{x^2}{2}\left(|\lambda_L|^2 + |\lambda_R|^{-2}\right)-x\,{\rm Im} \left(\lambda_L + \lambda_R^{-1}\right)}\,.
\end{equation}
This equation matches with eq. (16) of \cite{Chankowski:2004tb}.

\item {\it No $CP$ violation in mixing, i.e., $q/p$=1:} In this case, we have
\begin{itemize}
\item $|\lambda_L| = |\lambda_R|^{-1}$
\item $(\lambda_L + \lambda_R^{-1} )$ is real
\item ${\rm Im} (\lambda_L)= - {\rm Im} (\lambda_R^{-1})$
\item ${\rm Re} (\lambda_L)=  {\rm Re} (\lambda_R^{-1})$
\end{itemize}
With these simplifications, eq. \eqref{acp-lr} becomes 
\begin{equation}
A_{CP}^{LR}= - \frac{2\,x\,{\rm Im} (\lambda_L)\left((1+\lambda')^2+x^2\right)^{-1}}
{(1+|\lambda_L|^2+2\,y\,{\rm Re} (\lambda_L))(1-y^2)^{-1}+(1+\lambda')(1-|\lambda_L|^2)\left((1+\lambda')^2+x^2\right)^{-1}}\,.
\end{equation}

\item {\it No decoherence and no CP Violation in mixing i.e., $\lambda=0$, $\Delta \Gamma=0$ and $q/p=1$:} Under these assumptions, eq. \eqref{acp-lr} takes the very simple form
\begin{equation}
A_{CP}^{LR}= -\frac{2\,x\,{\rm Im} (\lambda_L)}{2+x^2+x^2\,|\lambda_L|^2}\,.
\end{equation}
This equation matches with eq. (19) of \cite{Chankowski:2004tb}.

\end{enumerate}

As previously indicated, the formalism outlined above is primarily suitable for the tau decay mode. In the SM, $Br(B_s^0\to\tau^+ \tau^-)$ is expected to be approximately 200 times greater than $Br(B_s^0 \to \mu^+  \mu^-)$ due to the significant mass difference between the tau and muon. Nevertheless, current experimental difficulties in tau reconstruction have limited us to obtaining only upper bounds on the branching ratio for this process. The LHCb collaboration provided the first direct constraint on the branching ratio of $B_s^0\to\tau^+ \tau^-$ in 2017 \cite{LHCb:2017myy} by using a dataset corresponding to an integrated luminosity of 3 $\rm {fb^{-1}}$. Assuming no contribution from $B_d^0\to\tau^+ \tau^-$ decays, the  upper limit for the branching fraction of $B_s^0\to\tau^+ \tau^-$ was obtained to be $6.8 \times 10^{-3}$ at a 95\% C.L.  Alternatively, assuming no contribution from $B_s^0\to\tau^+ \tau^-$ decay, an upper limit on $Br(B_d^0\to\tau^+ \tau^-)$ was obtained to be $2.1 \times 10^{-3}$ at a 95\% confidence level which is the most  stringent constraint  on $Br(B_d^0\to\tau^+ \tau^-)$. This limit is 2.6 times more stringent than the previous result from BaBar Collaboration \cite{BaBar:2005mbx}. 

The limit on the branching ratio is expected to be improved up to 2.6 to $5\times 10^{-4}$ at the end of Upgrade II of LHC \cite{LHCb:2018roe}. The expected upper limits on $Br(B_s^0\to\tau^+ \tau^-)$ and $Br(B_d^0\to\tau^+ \tau^-)$ that can be placed by the Belle Collaboration are of the order $10^{-4}$ and $10^{-5}$, respectively. Therefore, even with improved reconstruction techniques and data collection, the observation of these decay modes at the level of their SM values for the branching ratios is highly unlikely. Such detection would be possible at the  LHCb and Belle II only if new physics enhances its branching ratio up to the upper limits mentioned above. Such an enhancement may be feasible in a number of new physics models, see for e.g., \cite{Alonso:2015sja,Capdevila:2017iqn,Bordone:2023ybl}. Given the current analyses of futuristic sensitivities for these decays, it is obvious that neither LHC nor Belle II can measure these branching ratios up to the SM level. This is possible only at the FCC-ee which can not only measure its branching ratio at the level of SM but can also measure $CP$ structure  \cite{Bernardi:2022hny}.

On the other hand, the muonic decay mode shows promise from an experimental perspective. The decay process, specifically $B_s \to \mu^+ \mu^-$, has already been measured and future experiments can enhance its precision. To make effective use of this decay mode, we need to introduce additional observables that are not affected by the challenges in measuring muon polarization in the $B_s \to \mu^+ \mu^-$ decay mode. For this, we consider observables which are obtained by 
summing over the muon helicities.

We consider the following definitions:
\begin{equation}
    A_{\rm CP}^{\rm dir,\,h} = \frac{1-|\lambda_h|^2}{1+|\lambda_h|^2},  \quad A_{\Delta \Gamma}^{\rm h} = - \frac{2 {\rm Re}(\lambda_h)}{1+|\lambda_h|^2}, \quad A_{\rm CP}^{\rm mix,\,h} = -\frac{2 {\rm Im}(\lambda_h)}{1+|\lambda_h|^2}\,.
\end{equation}
Using these definitions and eq. \eqref{plep1},  we can write
\bea
\label{plep3}
\frac{P_{h}(B^0_s;t)}{\frac{1}{2} e^{-  \Gamma t}|A_h|^2} &=&
\left(1+|\lambda_h|^2\right) \Bigg[\cosh \left(\frac{ \Delta \Gamma t}{2}\right) 
+A_{\rm CP}^{\rm dir,\,h}  e^{-  \lambda t} \cos\left(\Delta m t\right) 
\nonumber\\
&&+ A_{\Delta \Gamma}^{\rm h} \sinh \left(\frac{ \Delta \Gamma t}{2}\right)
+A_{\rm CP}^{\rm mix,\,h}  e^{-  \lambda t} \sin\left(\Delta m t\right)\Bigg],\\
\label{plep4}
\frac{P_{h}(\bar{B}^0_s;t)}{\frac{1}{2} e^{-  \Gamma t}|A_h|^2 |\frac{p}{q}|^2} &=&
\left(1+|\lambda_h|^2\right) \Bigg[\cosh \left(\frac{ \Delta \Gamma t}{2}\right) 
-A_{\rm CP}^{\rm dir,\,h}  e^{-  \lambda t} \cos\left(\Delta m t\right) 
\nonumber\\
&&+ A_{\Delta \Gamma}^{\rm h} \sinh \left(\frac{ \Delta \Gamma t}{2}\right)
-A_{\rm CP}^{\rm mix,\,h}  e^{-  \lambda t} \sin\left(\Delta m t\right)\Bigg]\,.
\eea

Since it is difficult to measure the muon helicities, we consider the following rate
\begin{equation}
    P\left[B_s^0(t) \to \mu^+\, \mu^-  \right] = \sum_{h=L,R} P\left[B_s^0(t) \to \mu^+_{h}\, \mu^-_{h} \right] = P_{L}(B^0_s;t) + P_{R}(B^0_s;t)\,.
\end{equation}
We thus get
\begin{eqnarray}
  \frac{P\left[B_s^0(t) \to \mu^+\, \mu^-  \right]}{\frac{1}{2}e^{ -\Gamma t}} &=& \left[ |A_L|^2  \left(1+|\lambda_L|^2\right) + 
  |A_R|^2  \left(1+|\lambda_R|^2\right) \right] \cosh \left(\frac{ \Delta \Gamma t}{2}\right) 
  \nonumber\\
    &+& \left[ |A_L|^2  \left(1+|\lambda_L|^2\right) A_{\rm CP}^{\rm dir,\,L} + |A_R|^2  \left(1+|\lambda_R|^2\right) A_{\rm CP}^{\rm dir,\,R} \right] e^{-  \lambda t} \cos\left(\Delta m t\right) \nonumber\\
    &+& \left[ |A_L|^2  \left(1+|\lambda_L|^2\right) A_{\Delta \Gamma}^L + |A_R|^2  \left(1+|\lambda_R|^2\right) A_{\Delta \Gamma}^R \right] 
    \sinh \left(\frac{ \Delta \Gamma t}{2}\right)
    \nonumber\\
    &+& \left[ |A_L|^2  \left(1+|\lambda_L|^2\right) A_{\rm CP}^{\rm mix,\,L}  + |A_R|^2  \left(1+|\lambda_R|^2\right) A_{\rm CP}^{\rm mix,\,R} \right] e^{-  \lambda t} \sin\left(\Delta m t\right)\,.
\end{eqnarray}

Similarly, we get
\begin{eqnarray}
  \frac{P\left[\bar{B_s^0}(t) \to \mu^+\, \mu^-  \right]}{  \frac{1}{2}e^{ -\Gamma t} } &=& |\frac{p}{q}|^2 \left[|A_L|^2  \left(1+|\lambda_L|^2\right)+|A_R|^2  \left(1+|\lambda_R|^2\right) \right]\cosh \left(\frac{ \Delta \Gamma t}{2}\right) \nonumber\\
    &-& |\frac{p}{q}|^2 \left[|A_L|^2  \left(1+|\lambda_L|^2\right) A_{\rm CP}^{\rm dir,\,L} +|A_R|^2  \left(1+|\lambda_R|^2\right) A_{\rm CP}^{\rm dir,\,R} 
    \right] e^{-  \lambda t} \cos\left(\Delta m t\right) \nonumber\\
    &+& |\frac{p}{q}|^2 \left[ |A_L|^2  \left(1+|\lambda_L|^2\right) A_{\Delta \Gamma}^L + |A_R|^2  \left(1+|\lambda_R|^2\right) A_{\Delta \Gamma}^R
    \right] \sinh \left(\frac{ \Delta \Gamma t}{2}\right)\nonumber\\
    &-& |\frac{p}{q}|^2  \left[ |A_L|^2  \left(1+|\lambda_L|^2\right) A_{\rm CP}^{\rm mix,\,L}  + |A_R|^2  \left(1+|\lambda_R|^2\right) A_{\rm CP}^{\rm mix,\,R}
    \right]e^{-  \lambda t} \sin\left(\Delta m t\right)\,.
\end{eqnarray}
We can now define the following $CP$ asymmetry
\begin{equation}
    A_{\rm CP} (t) = \frac{P\left[B_s^0(t) \to \mu^+\, \mu^-  \right] - P\left[\bar{B_s^0}(t) \to \mu^+\, \mu^-  \right]}{P\left[B_s^0(t) \to \mu^+\, \mu^-  \right] +P\left[\bar{B_s^0} (t) \to \mu^+\, \mu^-  \right]}=\frac{N_{\rm CP}(t)}{D_{\rm CP}(t)}\,,
\end{equation}
where  $N_{\rm CP}(t)$ and $D_{\rm CP}(t)$ are  calculated to be
\begin{eqnarray}
    N_{\rm CP}(t) &=& e^{-\lambda t} \left[ A_{\rm CP}^{\rm dir}  \cos\left(\Delta m t\right)  + A_{\rm CP}^{\rm mix} \sin\left(\Delta m t\right)\right]
    - \delta_B \left[\cosh \left(\frac{ \Delta \Gamma t}{2}\right) + A_{\Delta \Gamma}\sinh \left(\frac{ \Delta \Gamma t}{2}\right)\right]\,,\nonumber\\
    D_{\rm CP}(t) &=& \cosh \left(\frac{ \Delta \Gamma t}{2}\right) + A_{\Delta \Gamma} \sinh \left(\frac{ \Delta \Gamma t}{2}\right) - \delta_B e^{-\lambda t}  \left[A_{\rm CP}^{\rm dir}\cos\left(\Delta m t\right)+ A_{\rm CP}^{\rm mix} \sin\left(\Delta m t\right) \right]\,,
\end{eqnarray}
with
\begin{eqnarray}
A_{\rm CP}^{\rm dir} &\equiv&  \frac{|A_L|^2 \left(1+|\lambda_L|^2\right) A_{\rm CP}^{\rm dir,\,L} + |A_R|^2 \left(1+|\lambda_R|^2\right) A_{\rm CP}^{\rm dir,\,R}}{|A_L|^2 \left(1+|\lambda_L|^2\right) + |A_R|^2 \left(1+|\lambda_R|^2\right)},\nonumber\\
A_{\rm CP}^{\rm mix} &\equiv& \frac{|A_L|^2 \left(1+|\lambda_L|^2\right) A_{\rm CP}^{\rm mix,\,L} + |A_R|^2 \left(1+|\lambda_R|^2\right) A_{\rm CP}^{\rm mix,\,R}}{|A_L|^2 \left(1+|\lambda_L|^2\right) + |A_R|^2 \left(1+|\lambda_R|^2\right)}\,,\nonumber\\
A_{\Delta \Gamma} &\equiv& \frac{|A_L|^2 \left(1+|\lambda_L|^2\right) A_{\Delta \Gamma}^{\rm L} + |A_R|^2 \left(1+|\lambda_R|^2\right) A_{\Delta \Gamma}^{\rm R}}{|A_L|^2 \left(1+|\lambda_L|^2\right) + |A_R|^2 \left(1+|\lambda_R|^2\right)}\,.
\end{eqnarray}
which simplifies to
\begin{eqnarray}
    A_{\rm CP}^{\rm dir} &=& \delta_B, \nonumber\\
    A_{\rm CP}^{\rm mix} &=& -\frac{2}{1+ |\frac{q}{p}|^2}\frac{|A_L|^2 \,{\rm Im}(\lambda_L)+|A_R|^2 \,{\rm Im}(\lambda_R)}{|A_L|^2 +|A_R|^2 }, \nonumber\\
    A_{\Delta \Gamma} &=& -\frac{2}{1+ |\frac{q}{p}|^2}\frac{|A_L|^2 \,{\rm Re}(\lambda_L)+|A_R|^2 \,{\rm Re}(\lambda_R)}{|A_L|^2 +|A_R|^2 }\,.
\end{eqnarray}
Thus we see that the helicity-independent $CP$ asymmetry also gets modulated by the presence of decoherence. As such symmetry can be measured by the current experimental facilities, it can be used to obtain bounds on the decoherence parameters. In the limit of neglecting the $CP$ violation in mixing, we get
\begin{equation}
    A_{\rm CP} (t) = \frac{ A_{\rm CP}^{\rm mix} \sin\left(\Delta m t\right)}{\cosh \left(\frac{ \Delta \Gamma t}{2}\right) +   A_{\Delta \Gamma} \sinh \left(\frac{ \Delta \Gamma t}{2}\right) } e^{-\lambda t}\,.
\end{equation}
Thus the dependence of time-dependent $CP$ asymmetry on the decoherence parameter survives even in the limit of neglecting the $CP$ violation in mixing. Given a tagging power of approximately 3.7\%, the availability of a dataset comprising 300 $\rm fb^{-1}$ enables the reconstruction of a pure sample encompassing over 100 flavor-tagged $B_s \to \mu^+\, \mu^-$ decays. This will facilitate the measurement of their time-dependent $CP$ asymmetry \cite{LHCb:2018roe}.

Let us now define the ``untagged'' rate:
\begin{equation}
    \left< P\left(B_s(t) \to \mu^+\, \mu^- \right) \right> \equiv 
    P\left(B_s^0(t) \to \mu^+\, \mu^- \right) +   P\left(\bar{B_s^0}(t) \to \mu^+\, \mu^- \right) \,.
\end{equation}
The calculation yields
\begin{eqnarray}
   \left< P\left(B_s(t) \to \mu^+\, \mu^- \right) \right> &=&
    \frac{1}{2}e^{ -\Gamma t}  \left(|A_L|^2 +|A_R|^2 \right) \left(2+|\frac{p}{q}|^2+|\frac{q}{p}|^2\right)   \nonumber\\&&  \times \Bigg[\cosh \left(\frac{ \Delta \Gamma t}{2}\right)+ A_{\Delta \Gamma} \sinh \left(\frac{ \Delta \Gamma t}{2}\right) 
  \nonumber\\&&   -\delta_B \left\{\delta_B \cos\left(\Delta m t\right) +  A_{\rm CP}^{\rm mix}\sin\left(\Delta m t\right) \right\}e^{-\lambda t}
    \Bigg]\,.
\end{eqnarray}
Thus it is evident that the untagged rate depends upon the decoherence parameter $\lambda$. When $|p/q|=|q/p|=1$, i.e. no CPV in mixing, then the above equation reduces to the following 
\begin{equation}
    \left< P\left(B_s(t) \to \mu^+\, \mu^- \right) \right> = 2 e^{ -\Gamma t}  \left(|A_L|^2 +|A_R|^2 \right) \left[\cosh \left(\frac{ \Delta \Gamma t}{2}\right)+ A_{\Delta \Gamma} \sinh \left(\frac{ \Delta \Gamma t}{2}\right) 
    \right].
   \end{equation}
which is independent of the decoherence parameter. 

The experimental branching ratio of $B_s \to \mu^+ \, \mu^-$ decay is given by
\begin{equation}
    Br(B_s \to \mu^+ \, \mu^-)_{\rm Exp} = \frac{1}{2}  
    \int_0^{\infty} \left< P\left(B_s(t) \to \mu^+\, \mu^- \right) \right> dt \,,
\end{equation}
which is calculated to be
\begin{eqnarray}
     Br(B_s \to \mu^+ \, \mu^-)_{\rm Exp} &=& \frac{1}{4\Gamma} \left(|A_L|^2 +|A_R|^2 \right) \left(2+|\frac{p}{q}|^2 + |\frac{q}{p}|^2\right) \Bigg[\frac{1}{1-y^2}+ A_{\Delta \Gamma} \frac{y}{1-y^2}\nonumber\\&&
     -\delta^2_B \frac{(1+\lambda')}{(1+\lambda')^2+x^2}-\delta_B \frac{x\, A_{\rm CP}^{\rm mix}}{(1+\lambda')^2+x^2}
     \Bigg]\,.
\end{eqnarray}
This quantity is modulated by the decoherence parameter. Now, the theoretical branching ratio is defined as
\begin{equation}
     Br(B_s \to \mu^+ \, \mu^-)_{\rm Theo} = \frac{1}{2\Gamma} \left< P\left(B_s(t) \to \mu^+\, \mu^- \right) \right> |_{\rm t=0}\,,
\end{equation}
which is calculated to be
\begin{equation}
    Br(B_s \to \mu^+ \, \mu^-)_{\rm Theo} = \frac{1}{4\Gamma} \left(|A_L|^2 +|A_R|^2 \right) \left(2+|\frac{p}{q}|^2 + |\frac{q}{p}|^2\right) (1-\delta_B^2)\,.
\end{equation}
We thus have,
\begin{equation}
     Br(B_s \to \mu^+ \, \mu^-)_{\rm Exp}  = \frac{Br(B_s \to \mu^+ \, \mu^-)_{\rm Theo} }{ (1-\delta_B^2)}
     \left[\frac{1+y  A_{\Delta \Gamma}}{1-y^2}-
     \frac{\delta_B^2(1+\lambda')+\delta_B\,x\,A_{\rm CP}^
     {\rm mix}}{(1+\lambda')^2+x^2}\right]\,.
\end{equation}
Thus we see that the relation between the theoretical and measured value of the branching ration of $B_s \to \mu^+ \, \mu^-$ is modulated by the presence of decoherence. However, in the limit of neglecting $CP$ violation in mixing, above relation reduces to 
\begin{equation}
     Br(B_s \to \mu^+ \, \mu^-)_{\rm Exp}  = Br(B_s \to \mu^+ \, \mu^-)_{\rm Theo}
     \left[\frac{1+y  A_{\Delta \Gamma}}{1-y^2}\right]\,,
\end{equation}
which is independent of decoherence parameter $\lambda$ and matches with the standard result obtained in \cite{DeBruyn:2012wk,Buras:2013uqa,Fleischer:2017ltw}.

We now consider the effective lifetime of the $B_s \to \mu^+ \, \mu^-$ decay, a parameter accessible through experimental observations of decay time distributions. These distributions are obtained from the same sets of untagged events utilized for measuring the branching ratios. The initial determination of the effective lifetime for the $B_s \to \mu^+\, \mu^-$ decay has been conducted by the LHCb collaboration, employing a dataset of 4.4 $\rm fb^{-1}$ and fitting the signal decay-time distribution. The measured effective lifetime is $2.04 \pm 0.44 \pm 0.05$ ps \cite{LHCb:2017rmj}. As more data becomes available, it is anticipated that the relative uncertainty on the effective lifetime will decrease to approximately 8\% with 23 $\rm fb^{-1}$ and further improve to 2\% with 300 $\rm fb^{-1}$ \cite{LHCb:2018roe}. The effective lifetime of the $B_s \to \mu^+ \, \mu^-$ decay  is theoretically defined as the time
the expectation value of the untagged rate given as
\begin{equation}
    \tau_{\mu^+ \, \mu^-} \equiv \frac{\int_0^{\infty}t \left< P\left(B_s(t) \to \mu^+ \, \mu^- \right) \right> dt}{\int_0^{\infty} \left< P\left(B_s(t) \to \mu^+ \, \mu^- \right) \right> dt}=\frac{1}{\Gamma}\frac{\tau_N}{\tau_D}\,,
\end{equation}
where $\tau_N$ and $\tau_D$  in the presence of decoherence is given by
\begin{eqnarray}
    \tau_N &=& \frac{1+y^2}{(1-y^2)^2}+A_{\Delta \Gamma}\frac{2y}{(1-y^2)^2}-\delta_B^2 \frac{(1+\lambda')^2-x^2}{[(1+\lambda')^2+x^2]^2}
-\delta_B A_{\rm CP}^{\rm mix}\frac{2x(1+\lambda')}{[(1+\lambda')^2+x^2]^2}\,,\\
\tau_D &=& \frac{1}{1-y^2}+A_{\Delta \Gamma}\frac{y}{1-y^2}-\delta_B^2 \frac{(1+\lambda')}{(1+\lambda')^2+x^2}   
-\delta_B A_{\rm CP}^{\rm mix}\frac{x}{(1+\lambda')^2+x^2}\,.
\end{eqnarray}
Thus we see that the decoherence parameter appears in the expression of the effective lifetime of $B_s \to \mu^+ \, \mu^-$ decay. However, in the limit $\delta_B \to 0$,  $\tau_{\mu^+ \, \mu^-}$ is given by
\begin{equation}
\tau_{\mu^+ \, \mu^-}=\frac{1}{(1-y^2)\Gamma} \frac{1+y^2+2yA_{\Delta \Gamma}}{1+yA_{\Delta \Gamma}}=\frac{\tau_{B_s}}{(1-y^2)} \frac{1+y^2+2yA_{\Delta \Gamma}}{1+yA_{\Delta \Gamma}}\,,
\end{equation}
which is independent of $\lambda$ and matches with the expression given in \cite{DeBruyn:2012wj}. Further, this limit allows the extraction of $y\,A_{\Delta \Gamma}$ through the measurement of the effective lifetime of 
$B_s \to \mu^+ \, \mu^-$:
\begin{equation}
y\,A_{\Delta \Gamma} =\frac{(1-y^2) \tau_{\mu^+ \, \mu^-} - \tau_{B_s}(1+y^2)}{2\tau_{B_s} - (1-y^2)\tau_{\mu^+ \, \mu^-}}\,.
\end{equation}
Thus, in the limit $\delta_B \to 0$, we get the following relation
\begin{equation}
  Br(B_s \to \mu^+ \, \mu^-)_{\rm Theo}  = \left[2-(1-y^2)\frac{\tau_{\mu^+ \, \mu^-} }{\tau_{B_s}}\right]Br(B_s \to \mu^+ \, \mu^-)_{\rm Exp}  \,,
\end{equation}
which is independent of the decoherence parameter. 

\section{Nonleptonic decays of $B$ mesons}
\label{nl}
In this section, we study the effect of decoherence on nonleptonic decays of $B$ mesons. Let us now consider $B^0/ \bar{B^0} \to f_{CP}$ decays where, 
$f_{CP}$ can be $J/\psi K_S$ or $D^+ D^-$ final states for $B^0_d$ and  $\psi \phi$ for $B^0_s$ meson. 
The operator for these decay modes can be written as
\be 
\mathcal{O}_{f_{CP}} = |A_{f}|^2\left(\begin{array}{ccc} 
1 & (\frac{p}{q})\,\lambda_{f} & 0  \\ (\frac{p}{q})^*\lambda_{f}^*  &|\frac{p}{q}|^2 |\lambda_f|^2 &0 \\ 0 & 0 & 0
\end{array}\right),
\label{opfcp}
\ee
where $\lambda_f$ is the phase invariant quantity defined as 
\be 
\lambda_f = \frac{q}{p}\frac{\bar{A_f}\left(\equiv A(\bar{B^0} \to f_{CP})\right)}{A_f \left(\equiv A(B^0 \to f_{CP})\right)}\,.
\ee
Therefore the probability rate that an initial state $B^0/ \bar{B^0}$ decays into final 
state $ f_{CP}$ is given by
\bea
\frac{P_{f_{CP}}(B^0;t)}{\frac{1}{2} e^{-  \Gamma t}|A_f|^2} &=&
\left(1+|\lambda_f|^2\right) \cosh \left(\frac{ \Delta \Gamma t}{2}\right) 
+\left(1-|\lambda_f|^2\right)e^{-  \lambda t} \cos\left(\Delta m t\right) 
\nonumber\\
&&- 2{\rm Re}(\lambda_f)\sinh \left(\frac{ \Delta \Gamma t}{2}\right)
-2 {\rm Im}(\lambda_f)e^{-  \lambda t} \sin\left(\Delta m t\right)\,,\nonumber\\
\frac{P_{f_{CP}}(\bar{B^0};t)}{\frac{1}{2} e^{-  \Gamma t}|A_f|^2 |\frac{p}{q}|^2} &=&
\left(1+|\lambda_f|^2\right) \cosh \left(\frac{ \Delta \Gamma t}{2}\right) 
-\left(1-|\lambda_f|^2\right)e^{-  \lambda t} \cos\left(\Delta m t\right) 
\nonumber\\
&&- 2{\rm Re}(\lambda_f)\sinh \left(\frac{ \Delta \Gamma t}{2}\right)
+2 {\rm Im}(\lambda_f)e^{- \lambda t } \sin\left(\Delta m t\right)\,.
\label{nonlep1}
\eea
One can then define a $CP$ violating observable
\be 
{\mathcal{A}}_{f_{CP}} (t) = \frac{P_{f_{CP}}(B^0;t)-P_{f_{CP}}(\bar{B^0};t)}
{P_{f_{CP}}(B^0;t)+P_{f_{CP}}(\bar{B^0};t)}.
\label{nonlep2}
\ee

Thus we have,
\be
{\mathcal{A}}_{f_{CP}} (t) = \frac{e^{-\lambda t}\left[A_{\rm CP}^{\rm dir,\,f_{CP}}   \cos\left(\Delta m t\right) + A_{\rm CP}^{{\rm mix},\,f_{CP}} \sin\left(\Delta m t\right) \right]-\delta_B\left[\cosh \left(\frac{ \Delta \Gamma t}{2}\right)-A_{\Delta \Gamma}^{f_{CP}}  \sinh \left(\frac{ \Delta \Gamma t}{2}\right)\right]}{\cosh \left(\frac{ \Delta \Gamma t}{2}\right)+A_{\Delta \Gamma}^{f_{CP}}  \sinh \left(\frac{ \Delta \Gamma t}{2}\right)- \delta_B e^{-\lambda t}\left[A_{\rm CP}^{\rm dir,\,f_{CP}}   \cos\left(\Delta m t\right) + A_{\rm CP}^{{\rm mix},\,f_{CP}}  \sin\left(\Delta m t\right) \right]}\,,
\ee
where
\begin{equation}
    A_{\rm CP}^{\rm dir,\,f_{CP}} = \frac{1-|\lambda_{f_{CP}}|^2}{1+|\lambda_{f_{CP}}|^2},  \quad A_{\Delta \Gamma}^{f_{CP}} = - \frac{2 {\rm Re}(\lambda_{f_{CP}})}{1+|\lambda_{f_{CP}}|^2}, \quad A_{\rm CP}^{{\rm mix},\,f_{CP}} = -\frac{2 {\rm Im}(\lambda_{f_{CP}})}{1+|\lambda_{f_{CP}}|^2},\,.
\end{equation}
Putting $\lambda=0$ in the above equation, we get the usual expression for $CP$ asymmetry in the interference of mixing and decay \cite{Waldi:2001rg}. These expressions with $\lambda=0$ in equations \eqref{nonlep1} and \eqref{nonlep2} are employed in all decay-time-dependent analyses of two-body B meson decays. Thus the presence of decoherence modifies the expression for $CP$ asymmetry in the interference of mixing and decay. Neglecting $CP$ violation in mixing, we get a simplified expression
\be 
{\mathcal{A}}_{f_{CP}} (t) = 
\frac{A_{\rm CP}^{\rm dir,\,f_{CP}}  \cos\left(\Delta m t\right)+ A_{\rm CP}^{{\rm mix},\,f_{CP}} \sin\left(\Delta m t\right)}
{ \cosh \left(\frac{ \Delta \Gamma t}{2}\right)+A_{\Delta \Gamma}^{f_{CP}}  \sinh \left(\frac{ \Delta \Gamma t}{2}\right)} e^{-  \lambda t} \,.
\label{cpasym1}
\ee
The time-dependent $CP$ asymmetry as given in eq. \eqref{cpasym1} with $\lambda=0$ is used to determine quantities like $\sin 2\beta$ and $\sin 2 \beta_s$ where $\beta \equiv \arg\left[ -(V^{\phantom{*}}_{cd}V_{cb}^*)/(V^{\phantom{*}}_{td}V_{tb}^*) \right]$ and $\beta_s \equiv \arg\left[ -(V^{\phantom{*}}_{ts}V_{tb}^*)/(V^{\phantom{*}}_{cs}V_{cb}^*) \right]$ are the angles of the unitarity triangle.  For e.g., the most statistically precise measurement of $\sin 2 \beta$ is obtained through a time-dependent angular analysis of $B^0_d \to J/\psi\, K_S$ which is also regarded as the golden channel to determine $\sin 2 \beta$. For this decay channel, $\Delta \Gamma=0$, $\lambda_f=1$ and in the limit of neglecting the penguin loop contributions to $b \to c \bar{c}s$ decay, we get
\begin{equation}
\mathcal{A}_{f_{CP}} (t) \approx \sin 2\beta  \cos\left(\Delta m t\right),
\end{equation}
which in the presence of decoherence is modulated by a factor of $e^{-\lambda t}$. Therefore the presence of decoherence may affect the clean determination of $\sin 2\beta$  and hence the standard time-dependent analysis of two body decays can be utilized to obtain bounds on the decoherence parameter.  The same is true for $\sin 2 \beta_s$ which is determined from the time-dependent angular analysis of $B^0_s \to J/\psi \,\phi$ decay mode. Similar analyses can be applied to determine these parameters using other decay modes. The decay channels like $B^0_d  \to \psi(2S)\, K_S$, $B^0_d \to J/\psi\, K^*$ and $B^0_d \to J/\psi \,\pi^0$ offer a means to extract $\sin 2 \beta$, while $\sin 2 \beta_s$ can also be extracted by performing time-dependent analyses on $B^0_s \to D_s^+\,D_s^-$, $B^0_s \to \psi(2S)\, \phi$ and $B^0_s \to J/\psi \, \eta^{(')}$. However, among these decay modes, the decays $B^0_d \to J/\psi \, K_S$ and $B^0_s \to J/\psi\, \phi$ yield the most precise determinations of $\sin 2 \beta$ and $\sin 2 \beta_s$ respectively. This is attributed not only to their accurate experimental measurements but also to their relatively cleaner theoretical predictions as compared to the other decay channels.

The SM predicts $\sin 2\beta$ from the CKM fit as $0.740^{+0.020}_{-0.025}$ \cite{Charles:2004jd, UTfit:2022hsi}. This prediction slightly conflicts with the current global average of $\sin \phi_d^{c\bar{c}s}=0.669 \pm 0.017$ \cite{HFLAV:2022esi}. Here, $\sin \phi_d^{c\bar{c}s}$ denotes the measured quantity, accommodating finite contributions from penguin loops. The prevailing global average of $\sin \phi_d^{c\bar{c}s}$ is principally governed by measurements from the BaBar, Belle, and LHCb collaborations in the $B^0_d \to J/\psi\, K_S$ decay. The most precise measurement comes from the Belle collaboration, yielding $0.670 \pm 0.029 \pm 0.013$ \cite{Belle:2012paq}, while the LHCb collaboration reports $0.731 \pm 0.035 \pm 0.020$ \cite{LHCb:2015ups}. It is apparent that the uncertainties in the LHCb and Belle measurements are comparable. This uncertainty is projected to decrease below 0.003 for the $B^0_d \to J/\psi\, K_S$ decay mode following the LHCb Upgrade II \cite{LHCb:2018roe}. Similarly, Belle II aims for a precision of about 0.006 after the accumulation of $50\, {\rm fb^{-1}}$ of data \cite{Belle-II:2018jsg}.

The quark level transition $b\to c\overline{c}s$ leads to the decay of $B_s^0 \to J/\psi \phi$ which is the potential decay mode to determine the value of $\sin 2\beta_s$. In the limit of neglecting penguin contributions to $b\to c\overline{c}s$ decay, $\phi_s^{c\bar{c}s}=-2\beta_s$. By performing a global fit to all relevant experimental data, the SM prediction of the phase $\phi_s^{c\bar{c}s}$ is obtained to be $-36.4\pm 1.2$ mrad \cite{Charles:2004jd} which is consistent with the experimental world average which is dominated by the LHCb measurement which in turn is dominated by the measurements involving time-dependent angular analysis of the $B_s^0 \to J/\psi (\mu^+\mu^-)\phi(K^+K^-)$ decay \cite{LHCb:2014iah}. After the upgrade II of LHCb, the precision on $\phi_s^{c\bar{c}s}$ is expected to be 4 mrad from $B_s^0 \to J/\psi \phi$ decay and 3 mrad from the combination of all modes. Thus the excellent precision on both channels provides exciting avenues to not only probe new physics but also provide much improved bounds on the current value of the decoherence parameter obtained from the flavor sector. 

It is crucial to highlight that leveraging measurements conducted in the precision era for probing sub-leading effects requires a thorough understanding and control of theoretical uncertainties in the decay channels mentioned above, stemming from hadronic effects. For instance, the quark-level transition $b\to c\bar{c}s$ is susceptible to ``penguin pollution" wherein, alongside tree-level contributions, additional contributions arise from gluonic and electroweak penguins. Consequently, reliable estimates of these higher-order corrections become imperative. Various methods have been proposed to assess hadronic corrections to $\phi_d^{c\bar{c}s}$ and $\phi_s^{c\bar{c}s}$, see for e.g. \cite{Faller:2008gt,Bhattacharya:2012ph,Fleischer:1999zi,Fleischer:2006rk,Jung:2012mp,DeBruyn:2014oga,Liu:2013nea,Frings:2015eva}. These methodologies have been put to the test through the examination of specific decay processes, such as $B^0_d \to J/\psi \rho^0$ \cite{LHCb-PAPER-2014-058} and $B_s^0 \to J/\psi \bar{K}^{*0}$ \cite{LHCb-PAPER-2015-034}. Currently, the most stringent constraint on penguin pollution in $\phi_s^{c\bar{c}s}$ is derived from the $B^0_d \to J/\psi \rho^0$ channel.

The time independent $CP$ asymmetry in the interference of mixing and decay, ${\mathcal{A}}_{f_{CP}}$, can be obtained by integrating the time dependent probability rates $P_{f_{CP}}(B^0(\bar{B^0});t)$ over time. We obtain
\begin{equation}
{\mathcal{A}}_{f_{CP}} =\frac{\left[A_{\rm CP}^{{\rm mix},\,f_{CP}} x +A_{\rm CP}^{\rm dir,\,f_{CP}}(1+\lambda')\right] \left[ (1+\lambda')^2+x^2\right]^{-1}-\delta_B \left(1-A_{\Delta \Gamma}^{f_{CP}} y\right)\left(1-y^2\right)^{-1}}{ \left(1+A_{\Delta \Gamma}^{f_{CP}}y\right)\left(1-y^2\right)^{-1} -\delta_B \left[A_{\rm CP}^{{\rm mix},\,f_{CP}} x +A_{\rm CP}^{\rm dir,\,f_{CP}}(1+\lambda')\right] \left[ (1+\lambda')^2+x^2\right]^{-1}}.
\end{equation}

In  the limit $\delta_B \rightarrow 0$, the above equation simplifies to
\begin{equation}
{\mathcal{A}}_{f_{CP}} =\frac{\left(1-y^2\right)\left[A_{\rm CP}^{{\rm mix},\,f_{CP}} x +A_{\rm CP}^{\rm dir,\,f_{CP}}(1+\lambda')\right] }{ \left(1+A_{\Delta \Gamma}^{f_{CP}} y\right) \left[ (1+\lambda')^2+x^2\right]}.
\end{equation}
The above integrated asymmetries can also be utilized to determine the value of the decoherence parameter by making use of the non-leptonic B decays.

\section{Conclusions}
\label{concl}
Flavor physics, particularly $B$ meson systems, offers an ingenious avenue for probing physics beyond the SM through precision measurements. The ongoing experiments at LHC and Belle II have already provided intriguing glimpses of potential new physics. Additionally, these experiments can also offer an opportunity to explore effects originating from much finer length scales. In this study, we focus on one such effect—quantum decoherence—that may emerge from Planck scale physics. Irrespective of the specific model responsible for generating decoherence, our analysis adopts a model-independent approach. We utilize the trace-preserving Kraus operator formalism where the interaction between the B meson system and its ambient environment is encoded by a single parameter. These operators extend the unitary evolution into a non-unitary framework while preserving entirely positive dynamics. This framework thus provides a versatile means to investigate the impact of quantum decoherence on B meson systems, allowing for a comprehensive exploration of potential effects arising from physics at much higher energies.

We investigate the impact of decoherence on a number of $B$ meson systems. These are meson anti-meson mixing, semileptonic decays of single and correlated $B$ mesons, and purely leptonic and non-leptonic decays of $B$ mesons. In these decay channels, we intend to identify observables which can be affected by the presence of decoherence effects. For all pertinent observables within these decay channels that are affected by decoherence, we provide theoretical expressions without neglecting the effects of $CP$ violating in mixing and decay width difference $\Delta \Gamma$.  Depending upon the theoretical and experimental precision of a specific decay mode, these effects can considered or ignored in an analysis. Our important findings are summarized as follows:

\begin{itemize}
    \item The decoherence parameter appears in the time-dependent mixing asymmetry which is used to determine mixing parameters $\Delta m$ and $\Delta \Gamma$. Notably, the effects of decoherence persist even when neglecting $\Delta \Gamma$ and assuming $|q/p|$ to be one. Consequently, a two-parameter (three-parameter) fit to this asymmetry can not only yield the true value of $\Delta m$ ($\Delta m$ and $\Delta \Gamma$) but also impose constraints on the decoherence parameter $\lambda$. 

    \item The semileptonic decays of a single B meson provide a number of asymmetries where decoherence parameter $\lambda$ appears. These asymmetries, both time-dependent as well as independent, can be precisely measured, offering a valuable avenue for determining the decoherence parameter.

    \item We also explore various time-dependent and time-independent observables in the decays of correlated $B$ mesons, exhibiting dependency on the decoherence parameter. Therefore these observables which include the ratio of the total same-sign to opposite-sign semileptonic rates as well as the total $B-\bar{B}$ mixing probability can serve as tools to establish constraints on the decoherence parameter $\lambda$. 

    \item The $CP$ asymmetries involving polarized decay probabilities in $B\to \ell \ell$ decays can also be used to determine $\lambda$. However, these observables can only be measured for the $\tau$ decay mode. 

    \item The decoherence parameter also appears in the lepton polarization independent $CP$ asymmetries in $B\to \ell \ell$ decays which can be measured for the muonic decay mode as well. The parameter $\lambda$ appears in the expression of the $CP$ asymmetry in the limit of neglecting $CP$ violation in mixing. 

    \item The decoherence parameter appears in the experimental branching ratio of $B_s \to \mu^+ \mu^-$ and hence in the well-known relation between the theoretical and experimental branching ratios of $B_s \to \mu^+ \mu^-$. However, in the limit of neglecting $CP$ violation in mixing, this relation remains unaffected by the decoherence effects.

    \item The theoretical expression of the effective lifetime for the $B_s \to \mu^+ \mu^-$ decay includes the decoherence parameter. However, the theoretical expression becomes independent of the parameter $\lambda$ if $q/p \to 1$.

    \item  The decoherence parameter also appears in time-dependent asymmetry associated with non-leptonic $B$ decays of type $B^0/ \bar{B^0} \to f_{CP}$. Hence the reanalysis of the determination of quantities such as $\sin 2 \beta$ and $\sin 2\beta_s$ in the presence of decoherence can provide tight limits. The time-independent asymmetry can also be used for obtaining such limits. 
\end{itemize}

Hence, we observe the presence of the decoherence parameter in various observables across all considered decay channels. In each of these channels, specific decay modes exhibit potential observables that are theoretically clean and can be determined with unparalleled precision. Consequently, the theoretical framework established in this study can become a valuable tool for deriving bounds on the decoherence parameters. This can be achieved by leveraging not only the forthcoming data from experiments in the near future but also through the reanalysis of existing data associated with numerous potential observables in the $B$ meson sector.

\acknowledgments
A. K. Alok would like to thank Subir Sarkar for useful suggestions and discussions. S. Uma Sankar acknowledges fruitful discussions  with Arantxa Oyenguren and Karim Trabelsi.  Additionally, A. K. Alok and S. Uma Sankar extend their thanks to the CERN Theory Division (CERN-TH) for their hospitality, where a part of this work was carried out.

\end{document}